\documentclass[usegraphicx,usenatbib]{mn2e}
\usepackage{times,amssymb}

\newcommand{\D}[2]{\frac{\partial #2}{\partial #1}}

\newcommand{\deriv}[2]{\frac{{\rm d} #2}{{\rm d}#1}}
\newcommand\bb[1]{\mbox{\boldmath{$#1$}}}
\newcommand\del{\bb{\nabla}}
\newcommand\bcdot{\bb{\cdot}}

\title[]
{A thermally stable heating mechanism for the intracluster medium: \\ turbulence, magnetic fields and plasma instabilities}
\author[Kunz et al.]
{M. W. Kunz$^{1}$\thanks{E-mail: kunz@thphys.ox.ac.uk}, A. A. Schekochihin$^{1}$, S. C. Cowley$^{2,3}$, J. J. Binney$^{1}$ and J. S. Sanders$^{4}$ \\
$^{1}$ Rudolf Peierls Centre for Theoretical Physics, University of Oxford, 1 Keble Road, Oxford, OX1 3NP, U. K.\\
$^{2}$ EURATOM/CCFE Fusion Association, Culham Science Centre, Abingdon, OX14 3DB, U. K.\\
$^{3}$ Blackett Laboratory, Imperial College, Prince Consort Road, London, SW7 2AZ, U. K.\\
$^{4}$ Institute of Astronomy, University of Cambridge, Madingley Road, Cambridge, CB3 0HA, U. K.
}
\date{Released 13/3/2010}
\pagerange{\pageref{firstpage}--\pageref{lastpage}} \pubyear{2010}
\def\LaTeX{L\kern-.36em\raise.3ex\hbox{a}\kern-.15em
    T\kern-.1667em\lower.7ex\hbox{E}\kern-.125emX}
\begin{document}
\label{firstpage} \maketitle

\begin{abstract}
We consider the problem of self-regulated heating and cooling in galaxy clusters and the implications for cluster magnetic fields and turbulence. Viscous heating of a weakly collisional magnetised plasma is regulated by the pressure anisotropy with respect to the local direction of the magnetic field. The intracluster medium is a high-beta plasma, where pressure anisotropies caused by the turbulent stresses and the consequent local changes in the magnetic field will trigger very fast microscale instabilities. We argue that the net effect of these instabilities will be to pin the pressure anisotropies at a marginal level, controlled by the plasma beta parameter. This gives rise to local heating rates that turn out to be comparable to the radiative cooling rates. Furthermore, we show that a balance between this heating and Bremsstrahlung cooling is thermally stable, unlike the often conjectured balance between cooling and thermal conduction. Given a sufficient (and probably self-regulating) supply of turbulent power, this provides a physical mechanism for mitigating cooling flows and preventing cluster core collapse. For observed density and temperature profiles, the assumed balance of viscous heating and radiative cooling allows us to predict magnetic-field strengths, turbulent velocities and turbulence scales as functions of distance from the centre. Specific predictions and comparisons with observations are given for several different clusters. Our predictions can be further tested by future observations of cluster magnetic fields and turbulent velocities.
\end{abstract}

\begin{keywords}
galaxies: clusters: intracluster medium -- magnetic fields -- instabilities -- turbulence -- diffusion -- conduction
\end{keywords}

\section{Introduction}

Early X-ray observations of galaxy clusters and the intracluster medium (ICM) indicated radiative losses large enough to lead to cooling flows \citep[see reviews by][]{sarazin86,sarazin88}. Mass deposition rates $\dot{M}_{\rm CF}$ due to presumed cooling flows were estimated to be as much as $\sim 10^3~{\rm M}_\odot~{\rm yr}^{-1}$ in some clusters \citep{cb77,fn77,mb78}. The cooling-flow model also predicted copious iron line emission from temperatures between $10^6$ and $10^7~{\rm K}$. However, at the time there was little direct evidence for mass dropout in any spectral band other than X-rays \citep[for a review, see][]{fabian94}. 

More recent high spectral resolution X-ray observations failed to detect the expected iron-line emission and constrained the central temperature to be $\sim 1/3$ of the bulk cluster temperature \citep{pf06}. The spectroscopically determined mass deposition rates are $\lesssim 0.1~\dot{M}_{\rm CF}$ \citep[e.g.][]{vf04}. This is despite the fact that the cooling time at $r\lesssim 100~{\rm kpc}$ is less than a Hubble time in $\gtrsim 70\%$ of clusters \citep[e.g.][]{esf92,pfeajw98,spo06,vkfjmmv06}. This discrepancy is the so-called `cooling flow problem.'

Some heating mechanism must therefore be balancing radiative cooling in the `cool-core' clusters. A wide variety of heating and heat transport schemes have been considered, including thermal energy from the outer regions of the cluster being transported to the central cooling gas by conduction \citep[e.g.][]{bc81,tr83,bm86,vsfaj02,fvm02,zn03}; energy in jets, bubbles, cosmic rays, outflows and/or radiation from a central active galactic nucleus (AGN) either resulting in turbulent diffusion of heat \citep[e.g.][]{dj96,kn03a} and/or being thermalised via dissipation of turbulent motions, sound waves and/or gravitational modes \citep[e.g.][]{lf90,bt95,co01,rbb04a,rbb04b,vd05,frtd05,brh05,cr07,mbbb08}; dynamical friction from galaxy wakes \citep[e.g.][]{schipper74,ly76,miller86,jdkbm90}; or some combination of these \citep[e.g.][]{brueggen03,dc05,co08}. Much of recent work has focused on either thermal conduction, AGN heating, or both. 

Thermal conduction alone cannot be the solution to the cooling flow problem across the full range of masses due to its steep temperature dependence \citep{vf04,kaastra04,ppkf06}. Even in hot systems where conduction is potent, fine tuning of the suppression factor $f$ (the fraction by which Spitzer conductivity is reduced due to, e.g., tangled magnetic field lines) is required \citep{bd88}. Moreover, if the thermal conduction has the same temperature dependence as the Spitzer conductivity (i.e. if $f$ is a constant) for a given ICM atmosphere, the resulting equilibria are thermally unstable (e.g. \citealt{bd88,soker03,kn03b}; see further discussion in Sections \ref{sec:balance} and \ref{sec:conduction}).

More recently, however, there has been renewed interest in the possibility that thermal conduction may provide sufficient heating to stably counteract the effects of radiative cooling. This has gone hand-in-hand with a dramatic increase in our understanding of `dilute' (i.e. only weakly collisional) plasmas, due to the appreciation that even very weak magnetic fields introduce an anisotropy into heat fluxes. One important consequence is that the criterion for convective instability changes to one of temperature, rather than entropy, increasing downwards \citep{balbus00,balbus01}. \citet{quataert08} generalised Balbus's (2000) analysis and found that a heat-flux buoyancy-driven instability (HBI) occurs for upwardly-increasing temperature profiles as well, so long as the magnetic field is not entirely horizontal (orthogonal to gravity). \citet{br08} conjectured that the nonlinear HBI is self-regulating and drives a reverse convective thermal flux, both of which may mediate the stabilisation of cooling cores. Numerical simulations of the HBI have been performed by \citet{pq08}, \citet{brbp09} and \citet{pqs09} with applications to the ICM. It was discovered that the HBI acts to rapidly reorient field lines to insulate the core, undermining the role of thermal conduction and causing a cooling catastrophe to occur. Subsequent work has shown that a moderate amount of turbulent driving may help regulate the HBI and allow thermal channels to remain open, potentially stabilising the core against collapse \citep{scqp09,pqs10,ro10}.

There are several reasons to believe that AGNs play an important role in regulating cooling. In many clusters, the AGN energy output inferred from radio-emitting plasma outflows and cavities is similar to the X-ray cooling rate of the central gas \citep{fabian00,ksh06,mn07,forman07}. Moreover, $\gtrsim 70\%$ of cool-core clusters harbour radio sources at their centres, while $\lesssim 25\%$ of clusters without cool cores are radio loud \citep{burns90}, providing strong circumstantial evidence for a connection between the processes that fuel the radio emission (such as AGNs) and the X-ray emission from the cooling gas. Models of self-regulated heating from AGN have been constructed \citep[e.g.][]{co01,rb02,bm03,kb03,hb04,go08,bs09} in which AGN activity is triggered by cooling-induced gas accretion toward cluster centres, increasing AGN heating and halting the collapse. Episodic outflows from AGN are thought not only to quench cooling and condensation in clusters, but also to limit the maximum luminosity of galaxies and regulate the growth of black holes at their centres \citep{binney05}.

While AGN activity is fundamentally linked with the observed presence of radio bubbles and/or X-ray cavities \citep[e.g.][]{brmwn04,df06}, it is currently unclear how the AGN energy is actually thermalised (e.g. see the introduction of \citealt{vd05} for a review of possibilities). This question can only be answered once knowledge of the effective viscosity of the ICM is acquired. The ICM hosts subsonic turbulence and magnetic fields with energy density comparable to that of the motions. Both of these should affect the viscosity of the ICM \citep[see review by][]{sc06}. In particular, the presence of a magnetic field alters the form of the viscosity when the ratio of the ion cyclotron and collision frequencies is much greater than unity \citep{braginskii65}, a condition amply satisfied in galaxy clusters. As a result, the transport properties of the ICM become strongly dependent on both the geometry and strength of the magnetic field, as well as on microscale plasma instabilities that are likely to occur ubiquitously in the ICM \citep[e.g. firehose and mirror;][]{sckhs05,lyutikov07,sckrh08,scrr10,rsrc10}.

In this paper, we investigate the effect these plasma effects might have on the large-scale transport properties of the ICM. Specifically, we argue that parallel viscous heating, due to the anisotropic damping of turbulent motions, is regulated by the saturation of microscale plasma instabilities (e.g., firehose and mirror) and can balance radiative cooling in the cool cores of galaxy clusters in such a way as to ensure thermal stability (Section \ref{sec:coolheat}). Given observed densities and temperatures, this balance implies specific values for central magnetic-field strengths and radial profiles of the rms magnetic field that are in good agreement with current observational estimates and that lend themselves to testing by future observations (Section \ref{sec:profiles}). We also show that, under the reasonable assumption that turbulent kinetic and magnetic energies are comparable to one another, cluster profiles for the turbulent velocity and the characteristic turbulence scale may be derived. The specific case of A1835 is considered as a typical example in Section \ref{sec:a1835}. Since the fundamental plasma-physical processes we appeal to are in principle universal in both cool-core and non-cool-core (i.e. unrelaxed) clusters, we also calculate in Section \ref{sec:coma} the magnetic-field strengths and turbulence characteristics for some non-cool-core clusters. They turn out to be in good agreement with current observational estimates. Thermal conduction and the robustness of our results with respect to it are briefly discussed in Section \ref{sec:conduction}. In Section \ref{sec:discussion}, we close with a discussion of our results and their limitations.

\section{Cooling and Heating of the ICM}\label{sec:coolheat}

\subsection{Radiative cooling}\label{sec:cool}

We follow \citet{tn01} in approximating the radiative cooling rate (per unit volume) of the ICM \citep[as determined by][]{sd93} by
\begin{equation}\label{eqn:cooling1}
Q^- = n_{\rm i} n_{\rm e} \Lambda(T) ,
\end{equation}
where the cooling function is
\begin{eqnarray}
\Lambda(T) &=& 10^{-23}~ {\rm erg~s}^{-1}~{\rm cm}^{3} \nonumber\\* &\times & \left[C_1\left(\frac{T}{1~{\rm keV}}\right)^{-1.7} + C_2\left(\frac{T}{1~{\rm keV}}\right)^{0.5} + C_3\right]  .
\end{eqnarray}
Our notation is standard: $n_{\rm i}$ ($n_{\rm e}$) is the number density of ions (electrons) and $T$ is the temperature (in energy units throughout the paper). We also define, for future use, the total number density $n = n_{\rm i} + n_{\rm e}$. The numerical constants $C_1 = 0.086$, $C_2=0.58$ and $C_3=0.63$ are selected to correspond to an average metallicity $Z=0.3~{\rm Z}_\odot$, for which the mean mass per particle is $\mu=0.597 m_{\rm p}$, the mean mass per electron is $\mu_{\rm e} = \mu (n/n_{\rm e}) = 1.150 m_{\rm p}$ and $n_{\rm i}=0.927 n_{\rm e}$ \citep{sd93}. The cooling function is dominated by Bremsstrahlung above $T\sim 1~{\rm keV}$ and by metal lines below $T\sim 1~{\rm keV}$. 

Since the temperature equilibration time between ions and electrons $t_{\rm i-e,eq}\sim 10~{\rm kyr}$ near the centres and $\sim 1~{\rm Myr}$ near the temperature maximum of cool-core clusters, which is smaller than all other timescales that will be relevant to us (see Section \ref{sec:conduction}), we assume $T_{\rm i}=T_{\rm e}=T$. Even in unrelaxed clusters like Coma, where the temperature is relatively high ($T\simeq 8.2~{\rm keV}$; \citealt{arnaud01}) and the electron density is relatively low ($n_{\rm e}\simeq 3\times 10^{-3}$ -- $4\times 10^{-5}~{\rm cm}^{-3}$; see Section \ref{sec:coma}), $t_{\rm i-e,eq}\simeq 2$ -- $170~{\rm Myr}$.

Normalised to conditions representative of the centres of cool-core clusters, the {\em radiative cooling rate} (per unit volume) is
\begin{eqnarray}\label{eqn:cooling2}
\lefteqn{Q^- \simeq 1.4\times 10^{-25} \left(\frac{n_{\rm e}}{0.1~{\rm cm}^{-3}}\right)^2 \left(\frac{T}{2~{\rm keV}}\right)^{1/2}~{\rm erg~s}^{-1}~{\rm cm}^{-3}} \nonumber\\*\mbox{}
\end{eqnarray}
in the Bremsstrahlung regime ($T\gtrsim 1~{\rm keV}$).

\subsection{Parallel viscous heating}\label{sec:heat}

There is a rapidly growing body of observational evidence for the presence of appreciable magnetic fields in the ICM \citep[for a review, see][]{ct02}. Randomly tangled magnetic fields with strength $B\sim 1$ -- $10~\mu{\rm G}$ and characteristic scale $\sim 1$ -- $10~{\rm kpc}$ are consistently found, with fields in the cool cores of cooling-flow clusters somewhat stronger than elsewhere.

The presence of a magnetic field alters the form of the thermal pressure when $\Omega_{\rm i}/\nu_{\rm ii}\gg 1$, where $\Omega_{\rm i}=eB/m_{\rm i}c$ is the ion cyclotron frequency and $\nu_{\rm ii}= 4\sqrt{\pi}n_{\rm i}\lambda_{\rm ii} e^4/3m^{1/2}_{\rm i}T^{3/2}$ is the ion-ion collision frequency; $\lambda_{\rm ii}$ is the ion-ion Coulomb logarithm \citep{braginskii65}. This is certainly the case in the cool cores of galaxy clusters, where typical values of the electron density $n_{\rm e}$, temperature $T$ and magnetic-field strength $B$ imply
\begin{equation}
\frac{\Omega_{\rm i}}{\nu_{\rm ii}} = 5.8\times 10^{10} \left(\frac{B}{10~\mu{\rm G}}\right) \left(\frac{n_{\rm e}}{0.1~{\rm cm}^{-3}}\right)^{-1} \left(\frac{T}{2~{\rm keV}}\right)^{3/2} .
\end{equation}
As a result, thermal pressure becomes anisotropic with respect to the local magnetic field direction $\bb{b}$:
\begin{equation}
\bb{\mathsf{P}} = p_\perp \left(\bb{\mathsf{I}}-\bb{b}\bb{b}\right) + p_{||}\bb{b}\bb{b} \equiv p~\bb{\mathsf{I}} + \bb{\sigma} ,
\end{equation}
where $p_\perp$ ($p_{||}$) is the thermal pressure perpendicular (parallel) to the local magnetic field, $p=(2/3)p_\perp + (1/3)p_{||}$ is the total thermal pressure, $\bb{\mathsf{I}}$ is the unit dyadic and we have defined the collisional viscous stress tensor
\begin{equation}\label{eqn:sigma}
\bb{\sigma} = -\left(\bb{b}\bb{b}-\frac{1}{3}\bb{\mathsf{I}}\right) \left(p_\perp - p_{||}\right) .
\end{equation}
In the Braginskii (i.e. collisional) limit, appropriate for the large-scale motions in the ICM since their dynamical timescales are $\gg \nu^{-1}_{\rm ii}\gg \Omega^{-1}_{\rm i}$, the ion contribution to the viscous stress dominates that of the electrons by a factor proportional to $(m_{\rm i}/m_{\rm e})^{1/2}$. Thus, in what follows, we neglect the electron contribution to the viscous stress.

The viscous stress tensor appears both in the momentum equation,
\begin{eqnarray}\label{eqn:momentum}
\lefteqn{m_{\rm i} n_{\rm i}\deriv{t}{\bb{u}} = -\del\bcdot\left[\left(p_{\rm i} + p_{\rm e} +\frac{B^2}{8\pi}\right)\bb{\mathsf{I}} -\frac{\bb{B}\bb{B}}{4\pi}+ \bb{\sigma}_{\rm i} \right] + m_{\rm i} n_{\rm i}\bb{g},}\nonumber\\*\mbox{}
\end{eqnarray}
as a form of momentum transport, and in the energy equation,
\begin{equation}\label{eqn:energy}
\frac{3}{2} n\deriv{t}{T} = -nT\del\bcdot\bb{u} - \bb{\sigma}_{\rm i}\bb{:}\del\bb{u} - \del\bcdot\bb{q}_{\rm e} - n_{\rm i}n_{\rm e}\Lambda(T) ,
\end{equation}
as a form of heating. In these two equations, $\bb{g}$ is the gravitational acceleration, $\bb{q}_{\rm e}$ is the electron collisional heat flux and ${\rm d}/{\rm d}t\equiv\partial/\partial t + \bb{u}\bcdot\del$ is the convective derivative. We assume an ideal gas equation of state, $p=nT$ (both for each species and for the two combined because $T_{\rm i}=T_{\rm e}=T$). The electron contribution to the collisional heat flux dominates that of the ions by a factor of $(m_{\rm i}/m_{\rm e})^{1/2}$ \citep{braginskii65}.

Differences between the perpendicular and parallel pressures in a magnetised plasma are due to the conservation of the first adiabatic invariant for each particle, $\mu = mv^2_\perp/2B = {\rm const}$ (on time scales $\gg \Omega_{\rm i}^{-1}$). Therefore, any change in the field strength must be accompanied by a corresponding change in the perpendicular pressure, $p_\perp/B\sim{\rm const}$. In a turbulent plasma such as the ICM, time-dependent fluctuations in the magnetic-field strength are inevitable. Accordingly, a patchwork of regions of positive or negative pressure anisotropy will emerge, corresponding to locally increasing or decreasing magnetic-field strength. If the pressure anisotropy $|p_\perp - p_{||}| \gtrsim B^2/4\pi$, firehose and mirror instabilities are triggered at spatial and temporal microscales \citep[and references therein]{sckhs05}.\footnote{In this paper we use the terms `microscales' and `microscopic' to describe processes whose lengthscales are just above the ion Larmor radius and whose growth rates are just below the ion cyclotron frequency.} Equations (\ref{eqn:momentum}) and (\ref{eqn:energy}) break down at these scales, and the perpendicular and parallel pressures must be determined by a kinetic calculation \citep[e.g.][]{scrr10,rsrc10}.

It is usually the case that the pressure anisotropy -- and thus the viscous stress -- is regulated by the nonlinear evolution of these microscale instabilities, which tend to pin the pressure anisotropy at marginal stability values \citep[see][and references therein]{rsrc10}:
\begin{equation}\label{eqn:marginality}
\Delta_{\rm i} \equiv \frac{p_{\perp,{\rm i}} - p_{||,{\rm i}}}{p_{\rm i}} = \frac{2\xi}{\beta_{\rm i}},
\end{equation}
where $\xi = -1$ for the firehose instability or $1/2$ for the mirror instability.\footnote{In a turbulent plasma, there will be regions of pressure anisotropy with both signs and so one might expect some average value of $\xi^2$ between $0.25$ and $1$. See also \citet{shqs06} for further refinements of this modelling.} The (ion) plasma beta parameter
\begin{equation}\label{eqn:beta}
\beta_{\rm i} \equiv \frac{8\pi n_{\rm i}T}{B^2} \simeq 75 \left(\frac{B}{10~\mu{\rm G}}\right)^{-2} \left(\frac{n_{\rm e}}{0.1~{\rm cm}^{-3}}\right) \left(\frac{T}{2~{\rm keV}}\right)
\end{equation}
is the ratio of the (ion) thermal and magnetic pressures, normalised here to conditions representative of the deep interiors of cool-core clusters. The recent observation that magnetic fluctuations in the solar wind are bounded by the firehose and mirror stability thresholds strongly supports the expectation that equation (\ref{eqn:marginality}) will be satisfied in a turbulent plasma \citep[e.g.][]{klg02,htkl06,bkhqss09}.

There are two fundamental physical ways in which marginal stability to microscale instabilities can be maintained in a weakly collisional plasma. {\em Either} microstabilities give rise to some effective particle scattering mechanism that isotropises the pressure \citep{sc06,shqs06,sqhs07} {\em or} they modify on the average the structure and time evolution of the magnetic field (or, equivalently, of the rate of strain) so as to cancel the pressure anisotropy caused by the changing fields. While there is no complete microphysical theory, existing calculations of particular cases \citep{sckrh08,chkpst08,ipb09,rsrc10,rsc10} all suggest the latter scenario. In this work we will {\em assume} that this is what happens. Importantly, this means that the collision rate is {\em not} modified by the instabilities. 

Let us now use this assumption to calculate the heating rate (per unit volume) due to parallel viscous dissipation of motions. From equations (\ref{eqn:sigma}) and (\ref{eqn:energy}), this rate is given by
\begin{equation}\label{eqn:heating1}
Q^+ = -\bb{\sigma}_{\rm i}\bb{:}\del\bb{u} = p_{\rm i}\Delta_{\rm i} \left(\bb{b}\bb{b}\bb{:}\del\bb{u} - \frac{1}{3}\del\bcdot\bb{u}\right) .
\end{equation}
We seek to write this equation solely in terms of the ion pressure anisotropy, and therefore we require an additional equation relating the rate of strain (the term in parentheses) to the ion pressure anisotropy. This is provided by \citet{braginskii65}:
\begin{equation}\label{eqn:braginskii}
\nu_{\rm ii}\Delta_{\rm i} = 2.9 \left(\bb{b}\bb{b}\bb{:}\del\bb{u} - \frac{1}{3}\del\bcdot\bb{u}\right) .
\end{equation}
This equation states that the rate of strain due to turbulent motions generates a pressure anisotropy that is relaxed on the ion-ion collision timescale. We interpret the right-hand side of equation (\ref{eqn:braginskii}) as including contributions to the (average) turbulent rate of strain from both the microscopic and macroscopic motions (see \citealt{rsrc10} for a detailed kinetic calculation of one example where this is justified). Under the assumption that the collision frequency is not modified by the instabilities, equation (\ref{eqn:braginskii}), together with equation (\ref{eqn:marginality}), states that microscale fluctuations adjust themselves in response to large-scale fluid motions in order to maintain and preserve marginality. 

Using equation (\ref{eqn:braginskii}) in equation (\ref{eqn:heating1}), we find that
\begin{equation}\label{eqn:heating2}
Q^+ = 0.35~p_{\rm i}\nu_{\rm ii}\Delta^2_{\rm i}.
\end{equation}
In other words, the ion pressure anisotropy is a source of free energy that is eventually converted into heat by collisions. With marginality to microscale instabilities imposed according to equation (\ref{eqn:marginality}), equation (\ref{eqn:heating2}) becomes
\begin{equation}\label{eqn:heating3}
Q^+ = 0.35~p_{\rm i}\nu_{\rm ii} \left(\frac{2\xi}{\beta_{\rm i}}\right)^2 = 2.2\times 10^{-3}~\xi^2 B^4 \frac{\nu_{\rm ii}}{p_{\rm i}}  .
\end{equation}
Normalised to conditions representative of the centres of cool-core clusters, this {\em parallel viscous heating rate} (per unit volume) is
\begin{eqnarray}\label{eqn:heating4}
\lefteqn{Q^+ = 10^{-25} ~\xi^2 \left(\frac{B}{10~\mu{\rm G}}\right)^4 \left(\frac{T}{2~{\rm keV}}\right)^{-5/2}~{\rm erg~s}^{-1}~{\rm cm}^{-3} .}\nonumber\\*\mbox{}
\end{eqnarray}
Note the lack of an explicit dependence on density and also, remarkably, on either the amplitude or the rate of strain of the turbulence. However, the strong dependence on $B$ does encode information about all of these. As the turbulent velocity increases, so too will the strength of the (dynamo-generated) magnetic field (see Section \ref{sec:velocities}), and the dissipation rate (eq. \ref{eqn:heating4}) will increase accordingly. In other words, the dissipation rate can be self-regulating. This local self-regulation is completely independent of whatever global self-regulation may be offered by the external source that is providing the turbulent energy (e.g. the AGNs), as long as there is enough turbulence to maintain the pressure anisotropy at its stability threshold (eq. \ref{eqn:marginality}). 

It should be stressed that the physical picture encapsulated by equation (\ref{eqn:heating4}) is somewhat different from the conventional approach to turbulence theory. Namely, in standard turbulence theory, one assumes that the energy input from external driving is fixed (i.e. given) and that all of that power is converted into turbulence and ultimately microphysically dissipated. In other words, the dissipation rate will adjust according to the rate at which external energy is input. In the ICM, there is a microscale constraint on the total turbulent rate of strain, namely, equation (\ref{eqn:marginality}). Although the dissipation can self-regulate via the strength of the magnetic field as described above, not all of the power provided by the external driving in fact need be locally thermalised via turbulence. Indeed, the turbulence may have an effective `impedance' and only accept the amount of power that can be locally viscously dissipated without triggering the microinstabilities. A further discussion of these points is given at the end of Section \ref{sec:discussion}.

\subsection{Thermal balance and stability}\label{sec:balance}

The similarity between the coefficients in equations (\ref{eqn:cooling2}) and (\ref{eqn:heating4}) is striking and strongly suggests that {\em parallel viscous heating can offset radiative losses in the deep interiors of cluster cores}. However, it is not enough simply to find a heating source capable of balancing cooling. One must also ensure that the resulting equilibrium is thermally stable. This can be shown for our proposed heating mechanism as follows. 

\begin{figure}
\centering
\includegraphics[width=3.2in]{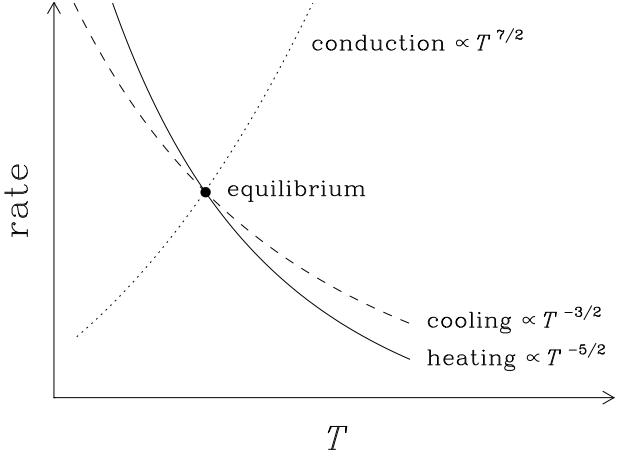}
\caption{A sketch of the rates of parallel viscous heating (solid line) and radiative cooling (dashed line) versus temperature $T$ at a fixed pressure. The thermal equilibrium point where heating balances cooling is denoted by a black dot. For a positive temperature perturbation from equilibrium, radiative cooling dominates over parallel viscous heating, whereas for a negative temperature perturbation from equilibrium, parallel viscous heating dominates over radiative cooling. Hence the equilibrium is stable to isobaric perturbations. By contrast, thermal balance between thermal conduction (dotted line) and radiative cooling is unstable.}
\label{fig:stability}
\end{figure}

In Fig.~\ref{fig:stability}, we show the rates of parallel viscous heating (solid line) and radiative cooling (dashed line) at fixed magnetic and thermal pressures as a function of temperature. The thermal equilibrium point where heating balances cooling is denoted by a black dot. The cooling rate (dashed line) scales as $T^{-3/2}$ in the Bremsstrahlung regime, while the parallel viscous heating rate (solid line) scales as $T^{-5/2}$. Hence, if the temperature is perturbed downwards, the heating rate increases faster than the rate of radiative cooling, and net heating restores the temperature to its equilibrium value. On the other hand, if the temperature is perturbed upwards, the heating rate decreases faster than the rate of radiative cooling, and net cooling restores the temperature to its equilibrium value. In other words, {\em parallel viscosity, regulated by the growth of microscale instabilities, endows the large-scale plasma with a source of viscous heating that makes the plasma thermally stable}.

This is in stark contrast to a thermal balance between radiative cooling and thermal conduction (dotted line), whose heating rate scales as $T^{7/2}$ at fixed thermal pressure. The resulting equilibrium is unstable, with an isobaric perturbation resulting in either an isothermal temperature profile (e.g. see fig.~7 of \citealt{pqs09}) or a cooling catastrophe. Thus, Spitzer conduction by itself cannot balance radiative cooling in a stable way \citep[e.g.][]{bd88,soker03,kn03b}.

The thermal stability argument for a balance between parallel viscous heating and radiative cooling may be refined by allowing for the possibility that the magnetic-field strength is not fixed, but rather depends on density and temperature in some unknown way. (For the remainder of this Section, we suppress the subscript ${\rm i}$ on $n$ and $\beta$ for economy of notation.) Again consider a thermal equilibrium, denoted by a subscript `$0$', in which energy losses balance energy gains:
\begin{equation}
Q^-(n_0,T_0) = Q^+(T_0,B_0) \equiv Q_0 ,
\end{equation}
where $Q^-$ is given by equation (\ref{eqn:cooling1}) and $Q^+$ by equation (\ref{eqn:heating3}). Allow small perturbations to this equilibrium, denoted by $\delta$, which preserve a constant {\em total} pressure $P$:
\begin{equation}\label{eqn:perturbedp}
\delta P = \delta p + \frac{B\delta B}{4\pi} + \frac{1}{3}\left(\delta p_\perp - \delta p_{||}\right) = 0 .
\end{equation}
This effectively removes sound waves from the analysis (see eq. \ref{eqn:momentum}) and allows us to focus on non-propagating, subsonic (`condensation') modes \citep[e.g.][]{field65}. With marginality to microscale instabilities imposed (eq. \ref{eqn:marginality}), equation (\ref{eqn:perturbedp}) becomes
\begin{equation}
\delta P = \delta p + \zeta\frac{B\delta B}{4\pi} = 0 ,
\end{equation}
where $\zeta \equiv 1 + 2\xi/3$. Put simply, the pressure anisotropy modifies the magnetic pressure. Then the perturbed cooling and heating rates are, respectively,
\begin{equation}\label{eqn:qmpert}
\left.\frac{\delta Q^-}{Q_0}\right|_{P} = 2\frac{\delta n}{n_0} + \frac{1}{2}\frac{\delta T}{T_0} ,
\end{equation}
\begin{equation}\label{eqn:qppert}
\left.\frac{\delta Q^+}{Q_0}\right|_{P} = 4\frac{\delta B}{B_0} - \frac{5}{2}\frac{\delta T}{T_0}.
\end{equation}
If we define
\begin{equation}
\theta_{T} \equiv \left.\D{\ln T}{\ln B}\right|_{n,0} \qquad {\rm and} \qquad \theta_{n} \equiv \left.\D{\ln n}{\ln B}\right|_{T,0},
\end{equation}
we find that the fractional perturbation in the density is related to the fractional perturbation in the temperature by
\begin{equation}\label{eqn:npert}
\frac{\delta n}{n_0} = -\frac{\delta T}{T_0} \frac{1+2\zeta\theta_{T}/\beta_0}{1+2\zeta\theta_{n}/\beta_0} .
\end{equation}
Note that this reduces to requiring a constant thermal pressure in the limit $\beta_0\gg 1$. Substituting equation (\ref{eqn:npert}) into equations (\ref{eqn:qmpert}) and (\ref{eqn:qppert}), we find
\begin{equation}
\left.\frac{\delta Q^-}{Q_0}\right|_{P} = +\frac{\delta T}{T_0} \left(\frac{1}{2} - 2~\frac{1+2\zeta\theta_{T}/\beta_0}{1+2\zeta\theta_{n}/\beta_0}\right) ,
\end{equation}
\begin{equation}
\left.\frac{\delta Q^+}{Q_0}\right|_{P} = -\frac{\delta T}{T_0}\left(\frac{5}{2} - 4~\frac{\theta_{T}-\theta_{n}}{1+2\zeta\theta_{n}/\beta_0}\right).
\end{equation}
From this, it follows that the loss function (i.e. heat losses minus heat gains) is
\begin{eqnarray}
\left.\mathcal{L}(n,T,B)\right|_{P} &\equiv & \left.Q^-(n,T,B)\right|_{P} - \left.Q^+(n,T,B)\right|_{P} \nonumber\\*\mbox{} &=& Q_0 \frac{\delta T}{T_0} \left[1 - 4\left(1+\frac{\zeta}{\beta_0}\right)\left(\frac{\theta_{T}-\theta_{n}}{1+2\zeta\theta_{n}/\beta_0}\right) \right].\nonumber\\*\mbox{}
\end{eqnarray}
Since $\beta\gg 1$ in the ICM (see eq. \ref{eqn:beta}),
\begin{equation}\label{eqn:loss}
\left.\mathcal{L}(n,T,B)\right|_{P} \simeq Q_0 \frac{\delta T}{T_0}\Bigl[ 1 - 4 \bigl( \theta_{T} - \theta_{n} \bigr) \Bigr] .
\end{equation}
To ensure thermal stability, the term in square brackets must be non-negative, so that the loss function has the same sign as $\delta T$. Then, by equation (\ref{eqn:loss}),
\begin{equation}\label{eqn:stability}
\left.\D{\ln T}{\ln B}\right|_{n} - \left.\D{\ln n}{\ln B}\right|_{T} \le \frac{1}{4}  \quad \textrm{(for thermal stability).}
\end{equation}
This inequality is marginally satisfied (by definition) if heating balances cooling. Unless the perturbed magnetic field is a strongly increasing function of temperature or the magnetic field increases with decreasing density, neither of which are particularly likely, the inequality (\ref{eqn:stability}) is satisfied and so the equilibrium is thermally stable.

\section{Implications: cluster equilibrium profiles}\label{sec:profiles}

Given the apparent ability of parallel viscous heating to balance radiative cooling and stabilise the centres of cool-core clusters, it is tempting to ascribe the entire temperature profile inside the cooling radius to a long-term balance between this form of heating and cooling. In this Section, we investigate what this implies for magnetic-field strengths, turbulent velocities and turbulence scales throughout cluster cores as functions of radius $r$. We then assess {\it a posteriori} whether or not the results are observationally permissible and theoretically sensible. 

\subsection{Magnetic fields}\label{sec:bfields}

\begin{table}
\centering
\caption{Predicted central magnetic-field strengths $B_{\rm c,theory}$ for a variety of clusters. Central electron number densities $n_{\rm e,c}$ and temperatures $T_{\rm c}$ for each cluster are taken from (A1835) \citet{sfsp10}; (Hydra A) \citet{dnmfjprw01}; (A478, A1795) \citet{dc05}; (A2199) \citet{jafs02}; (M87) \citet{gmpd04}; (Centaurus, A262) \citet{cdvs09}; (A2142, A401) \citet{markevitch98}, \citet{crbiz07}; (Ophiuchus, A2634, A400) \citet{crbiz07}, \citet{fmt98}; (A2382) \citet{evb96}, \citet{gmgpgrcf08}; and (A2255) \citet{fbgn97}, \citet{gmfgdt06}, \citet{dw98}. Observationally-inferred magnetic-field strengths $B_{\rm c,obs}$, where available, are taken from (Hydra A) \citet{ve03}; (A2199, M87, A1795, A2634, A400) \citet{eo02}; (Centaurus) \citet{tfa02}; (A2382) \citet{gmgpgrcf08}; and (A2255) \citet{gmfgdt06}. The non-cool-core cluster results are discussed in Section \ref{sec:coma}.}
\begin{minipage}{3.21in}
\centering
\renewcommand{\thefootnote}{\thempfootnote}
\begin{tabular}{lcccc}
\hline \hline
			& $n_{\rm e,c}$ 			& $T_{\rm c}$	& $B_{\rm c,theory}$		& $B_{\rm c,obs}$ \\
Cluster name	& ($10^{-2}~{\rm cm}^{-3}$)	& (keV)		& ($\xi^{-1/2}\mu{\rm G}$)	& ($\mu{\rm G}$) \\
\hline
\multicolumn{5}{c}{Cool-core clusters}\\
\hline
A1835 & 10 & 2.85 & 13.8 & -- \\
Hydra A & 7.2 & 3.11 & 12.4 & 12$^a$ \\
A478 & 15.2 & 1.72 & 12.1 & -- \\
A2199 & 10 & $\simeq 2$ & $\simeq 11$ & 15$^b$ \\
M87 & 10.8 & 1.62 & 9.8 & 35$^b$ \\
A1795 & 5.4 & 2.26 & 8.6 & 9.7$^b$ \\
Centaurus & 9.5 & 1.24 & 7.7 & $8$ \\
A262 & 3.7 & 1.54 & 5.5 & -- \\
\hline
\multicolumn{5}{c}{Non-cool-core clusters}\\
\hline
A2142 & 1.87 & 8.8 & 13.0 & RM$^c$ \\
Ophiucus & 0.80 & 10.3 & 9.5 & RM$^c$ \\
A401 & 0.70 & 8.3 & 7.6 & RM$^c$ \\
A2382 & 0.50 & 2.9 & 3.1 & $3$ \\
A2634 & 0.28 & 3.7 & 2.7 & 3.5$^b$ \\
A2255 & 0.2 & 3.5 & 2.2 & $2.5$ \\
A400 & 0.24 & 2.3 & 1.8 & 2.9$^b$ \\
\hline
\end{tabular}
\footnotetext[1]{Estimates for $B_{\rm c,obs}$ in Hydra A are widely varying, from $7~\mu{\rm G}$ \citep{ve05} and $12~\mu{\rm G}$ \citep{ve03}, all the way up to $\sim 30~\mu{\rm G}$ \citep{tp93,ke09}.}
\footnotetext[2]{For these clusters, $B_{\rm c,obs}$ is inferred by assuming that only one magnetic filament along the line of sight accounts for the observed rotation measure; therefore these are likely upper limits.}
\footnotetext[3]{`RM' indicates that a rotation measure exists, but that no $B_{\rm c,obs}$ has been inferred \citep[see][and references therein]{gdm10}.}
\label{tab:bfields}
\end{minipage}
\end{table}

We assume that parallel viscous heating (eq. \ref{eqn:heating4}) due to turbulent dissipation balances radiative cooling (eq. \ref{eqn:cooling2}) at all radii inside the cluster core:
\begin{equation}\label{eqn:heatingiscooling}
Q^+(r) \simeq Q^-(r) .
\end{equation}
In the Bremsstrahlung regime, this implies
\begin{equation}\label{eqn:Bprofile}
B \simeq 11 ~\xi^{-1/2} \left(\frac{n_{\rm e}}{0.1~{\rm cm}^{-3}}\right)^{1/2}\left(\frac{T}{2~{\rm keV}}\right)^{3/4} ~\mu{\rm G}.
\end{equation}
Note that {\em the predicted magnetic-field strength is thus a function of density and temperature}. It is common in both numerical modelling of clusters \citep[e.g.][]{dsgf01} and in data analysis aiming to reconstruct magnetic-field strengths and spectra \citep[e.g.][]{mgfgdftd04,ke09} to assume an exclusive relationship between $B$ and $n_{\rm e}$. Our arguments suggest that these models may need to be generalised to accommodate the temperature dependence.

For typical electron densities ($n_{\rm e}\sim 0.01$ -- $0.1~{\rm cm}^{-3}$) and temperatures ($T\sim 1$ -- $3~{\rm keV}$) at the centres of cool-core clusters, equation (\ref{eqn:Bprofile}) implies central magnetic fields $\sim 1$ -- $10~\mu{\rm G}$, within observational constraints. For example, conditions near the centre of the popular Hydra A cluster ($n_{\rm e,c}\simeq 0.072~{\rm cm}^{-3}$ and $T_{\rm c}\simeq 3.11~{\rm keV}$; \citealt{dnmfjprw01}; H. Russell, private communication) imply a thermal-equilibrium magnetic-field strength $B_{\rm c}\simeq 12.4~ \xi^{-1/2}~\mu{\rm G}$. Farther out around $\simeq 30~{\rm kpc}$, the observed electron density $n_{\rm e}\simeq 0.02~{\rm cm}^{-3}$ and temperature $T\simeq 3.5~{\rm keV}$ imply $B\simeq 7~ \xi^{-1/2}~\mu{\rm G}$. These are both in good agreement with magnetic-field strength estimates in Hydra A from Faraday rotation maps \citep{ve03,ve05}. For another cluster, A2199, popular among theorists \citep[e.g.][]{pqs09}, a central density $n_{\rm e,c} \simeq 0.1~{\rm cm}^{-3}$ and central temperature $T_{\rm c}\simeq 2~{\rm keV}$ (\citealt{jafs02}; H. Russell, private communication) imply $B_{\rm c}\simeq 11~\xi^{-1/2}~\mu{\rm G}$. \citet{eo02} inferred a central magnetic-field strength there of $15~\mu{\rm G}$ by assuming that only one magnetic filament along the line of sight accounts for the observed rotation measure. In Table \ref{tab:bfields}, we list these and other central magnetic-field strength predictions.

\subsection{Turbulent velocities}\label{sec:velocities}

In order to estimate the turbulent velocities in the ICM, we assume that the large-scale kinetic and magnetic energies are in overall equipartition, 
\begin{equation}\label{eqn:equipartition}
\frac{1}{2} m_{\rm i} n_{\rm i} U^2_{\rm rms} \simeq \frac{B^2}{8\pi} ,
\end{equation}
where $U_{\rm rms}\equiv \langle u^2\rangle^{1/2}$ is the rms flow velocity. This is expected to be the case for a magnetic field amplified and brought to saturation by the fluctuation dynamo \citep[see, e.g., review by][and references therein]{sc07}. Then the rms turbulent velocity is equal to the Alfv\'{e}n speed and, using equation (\ref{eqn:Bprofile}), we therefore obtain
\begin{equation}\label{eqn:Uprofile}
U_{\rm rms} \simeq 70 ~\xi^{-1/2}\left(\frac{T}{2~{\rm keV}}\right)^{3/4}~{\rm km~s}^{-1}
\end{equation}
in the Bremsstrahlung regime. In other words, {\em relatively hotter (cooler) cores should therefore have larger (smaller) turbulent velocities}. The corresponding Mach number is
\begin{equation}
M \equiv \frac{U_{\rm rms}}{c_{\rm s}} = 0.18 ~\xi^{-1/2}\left(\frac{T}{2~{\rm keV}}\right)^{1/4},
\end{equation}
where $c_{\rm s} = (T/m_{\rm i})^{1/2}$ is the sound speed. Note the weak dependence of $M$ on temperature. For $T\simeq 1$ -- $10~{\rm keV}$, $U_{\rm rms}$ ranges from $\simeq 44~\xi^{-1/2}~{\rm km~s}^{-1}$ to $210~\xi^{-1/2}~{\rm km~s}^{-1}$, so $M$ ranges from $\simeq 0.16~\xi^{-1/2}$ to $0.24~\xi^{-1/2}$. While the turbulent velocities are not yet measured directly, theoretical \citep[e.g.][]{dc05,ev06,ssh06}, numerical \citep[e.g.][]{nb99,rs01,snb03} and indirect observational \citep[e.g.][]{sfmbb04,cfjsb04,rcbf05,rcbf06,gfsm06,rcsbf08,sfs10} estimates suggest that the numbers we are predicting are reasonable.

We caution here that equation (\ref{eqn:Uprofile}) should be considered a lower limit on the actual turbulent velocity since $U_{\rm rms}$ is unlikely to be smaller, but could be larger, than the Alfv\'{e}n speed. In magnetohydrodynamic numerical simulations, it is often larger by a factor of order unity \citep[e.g.][]{sctmm04,hbd04}.

\begin{figure*}
\centering
\includegraphics[width=2.3in]{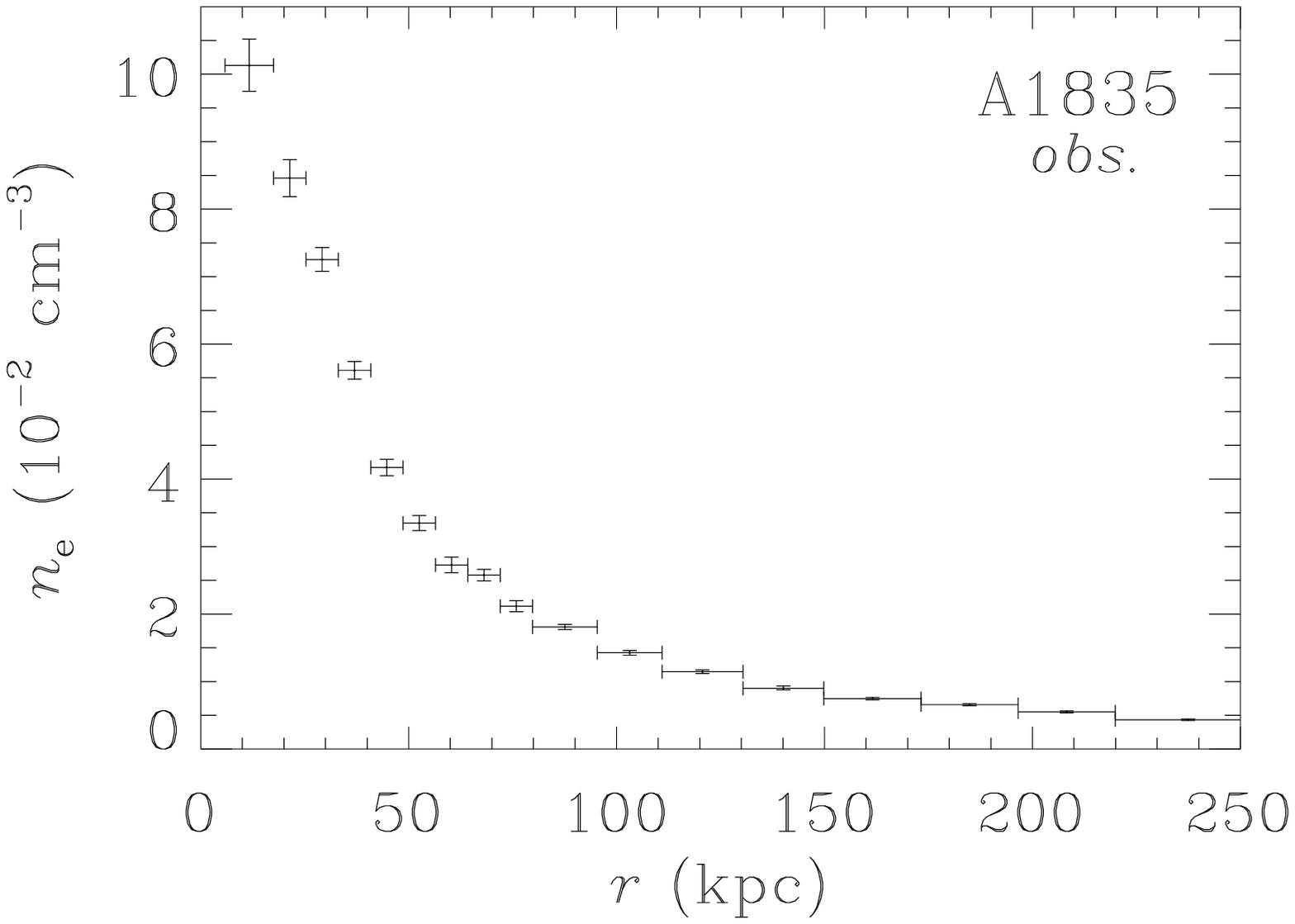}
\includegraphics[width=2.3in]{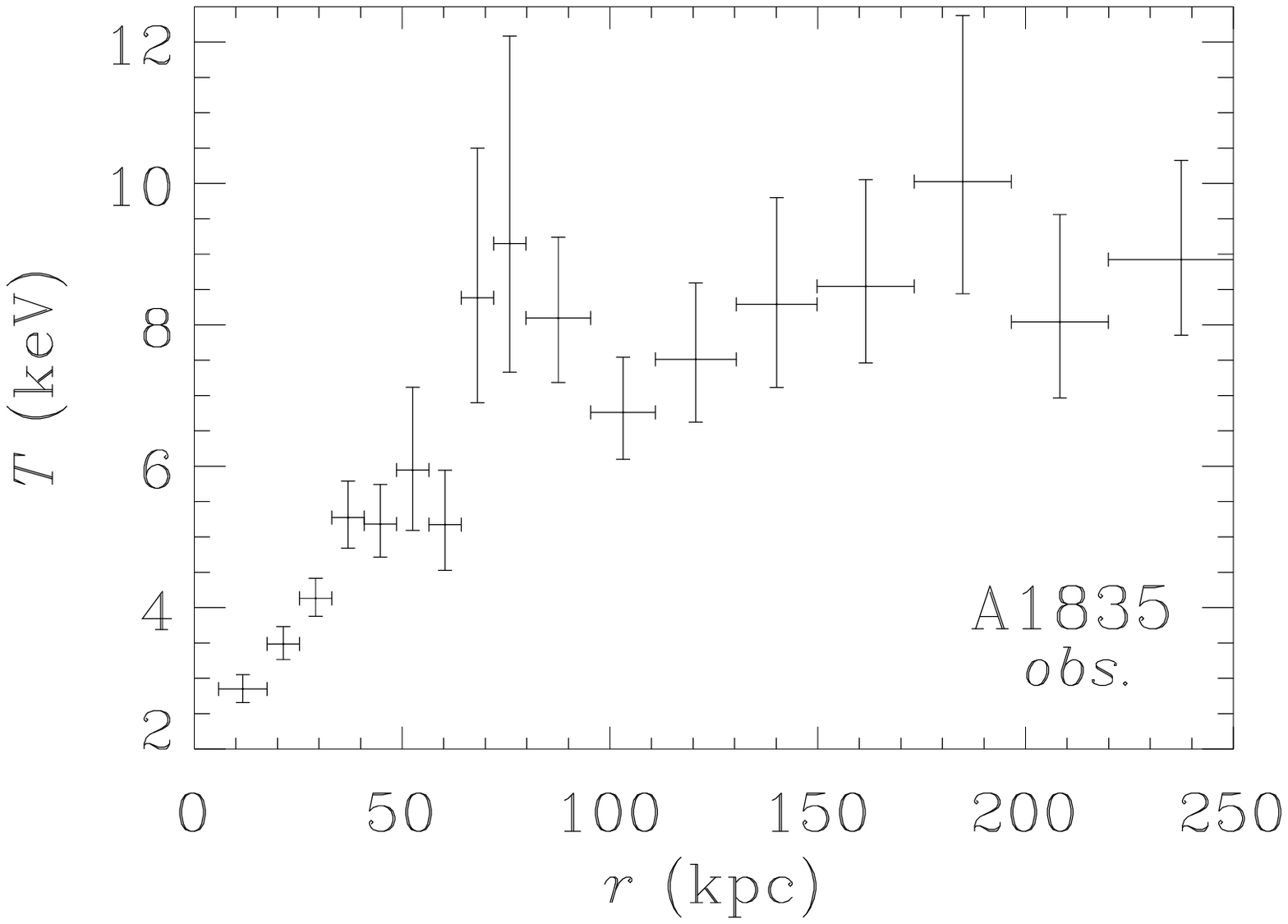}
\includegraphics[width=2.3in]{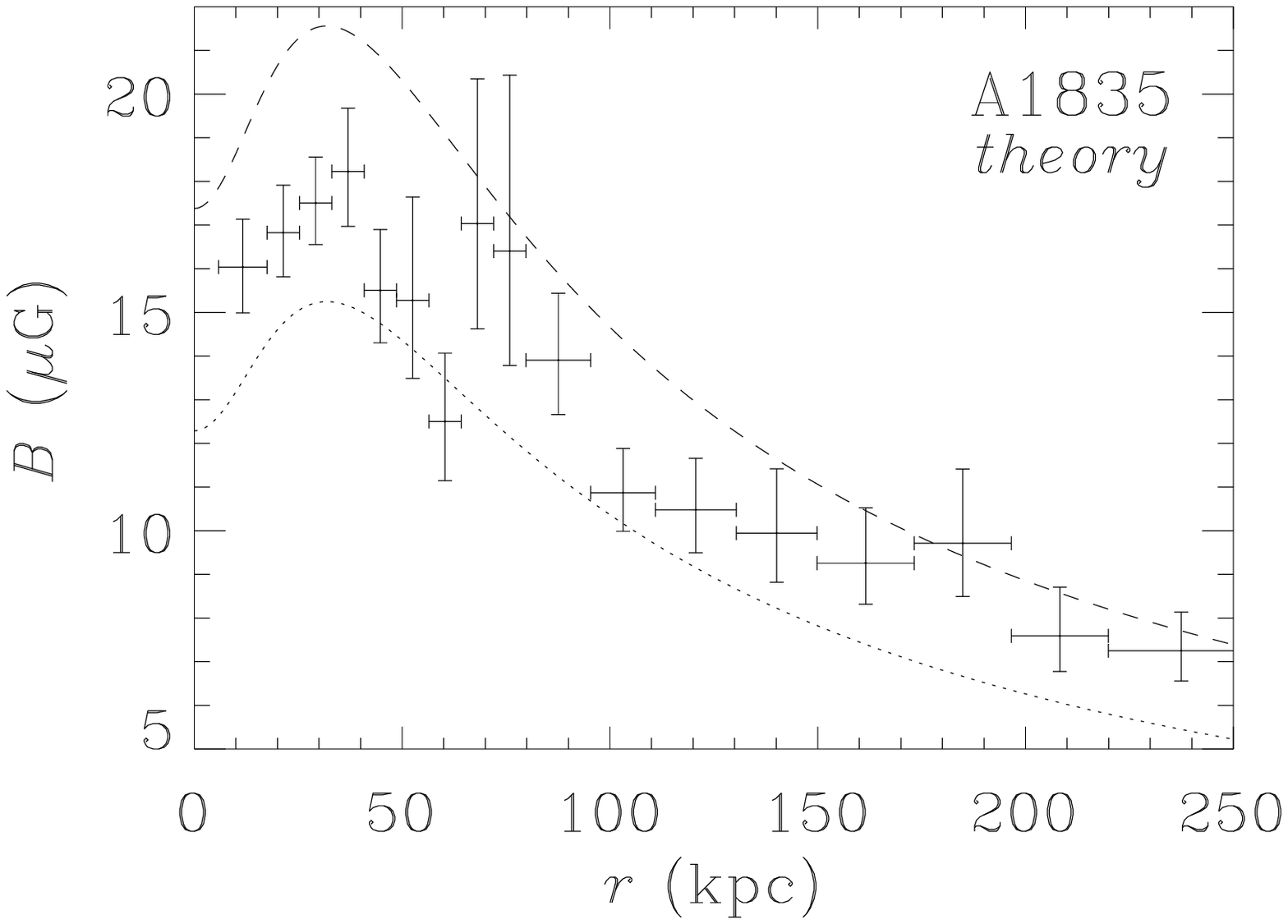}
\newline
\includegraphics[width=2.3in]{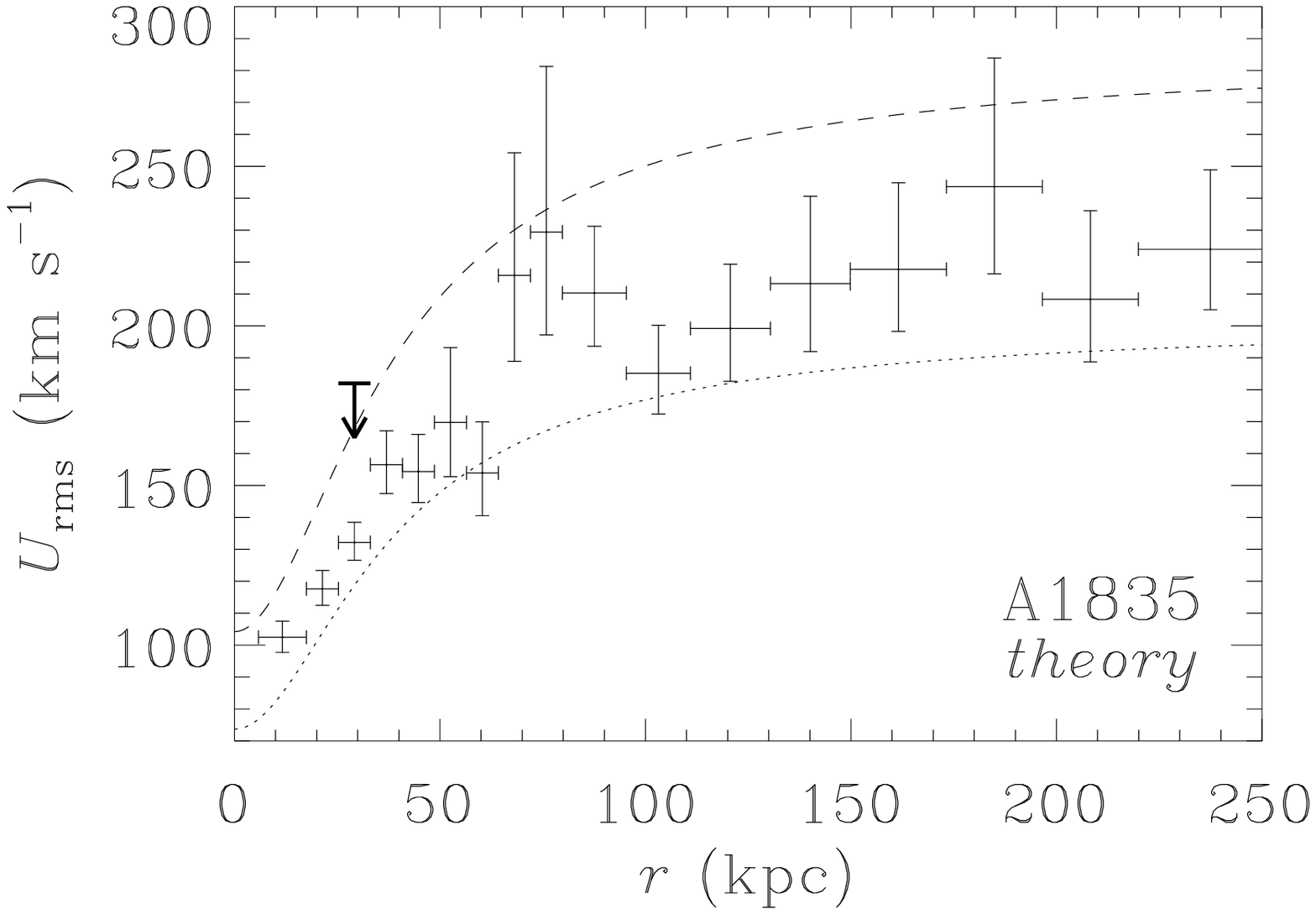}
\includegraphics[width=2.3in]{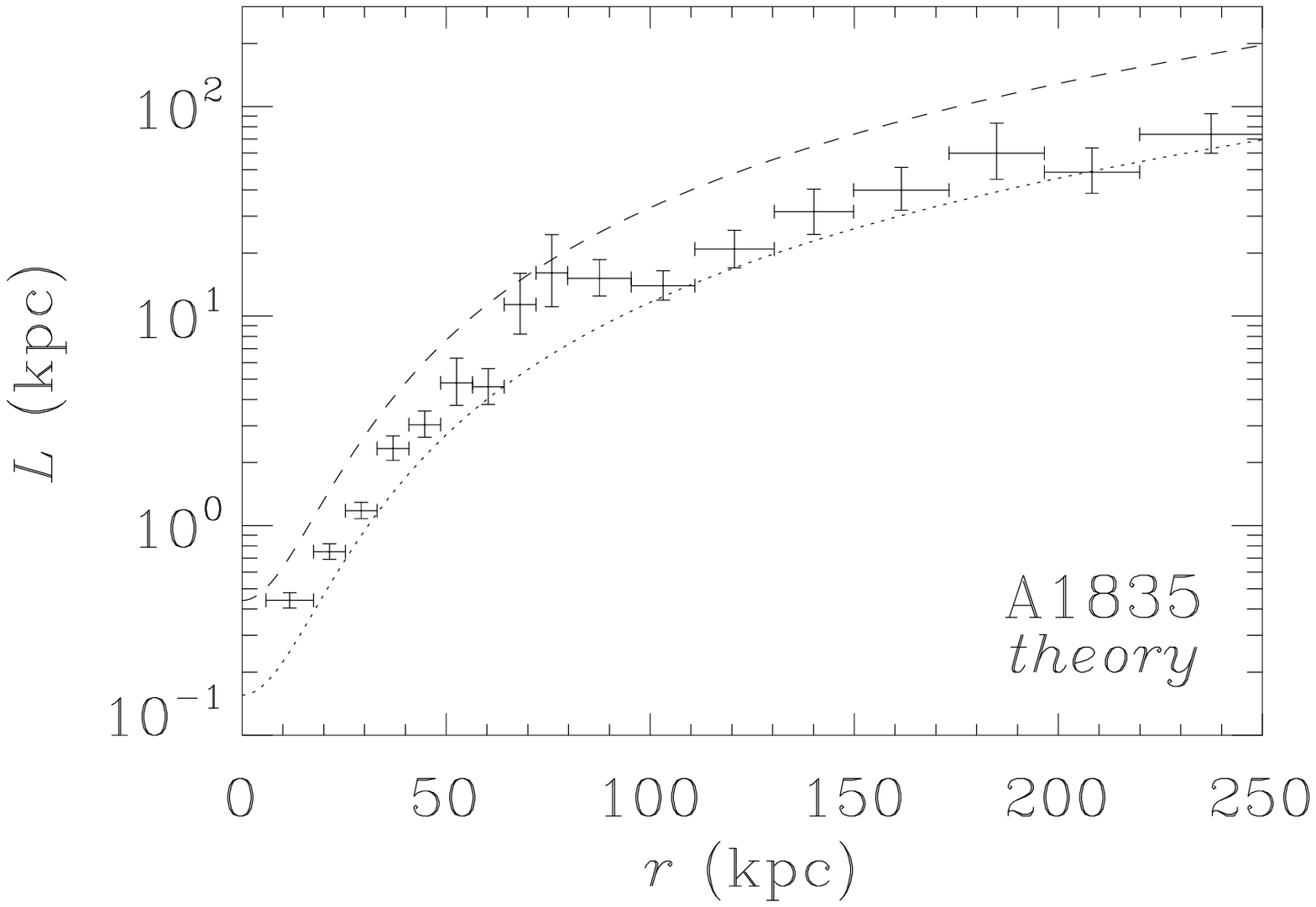}
\includegraphics[width=2.3in]{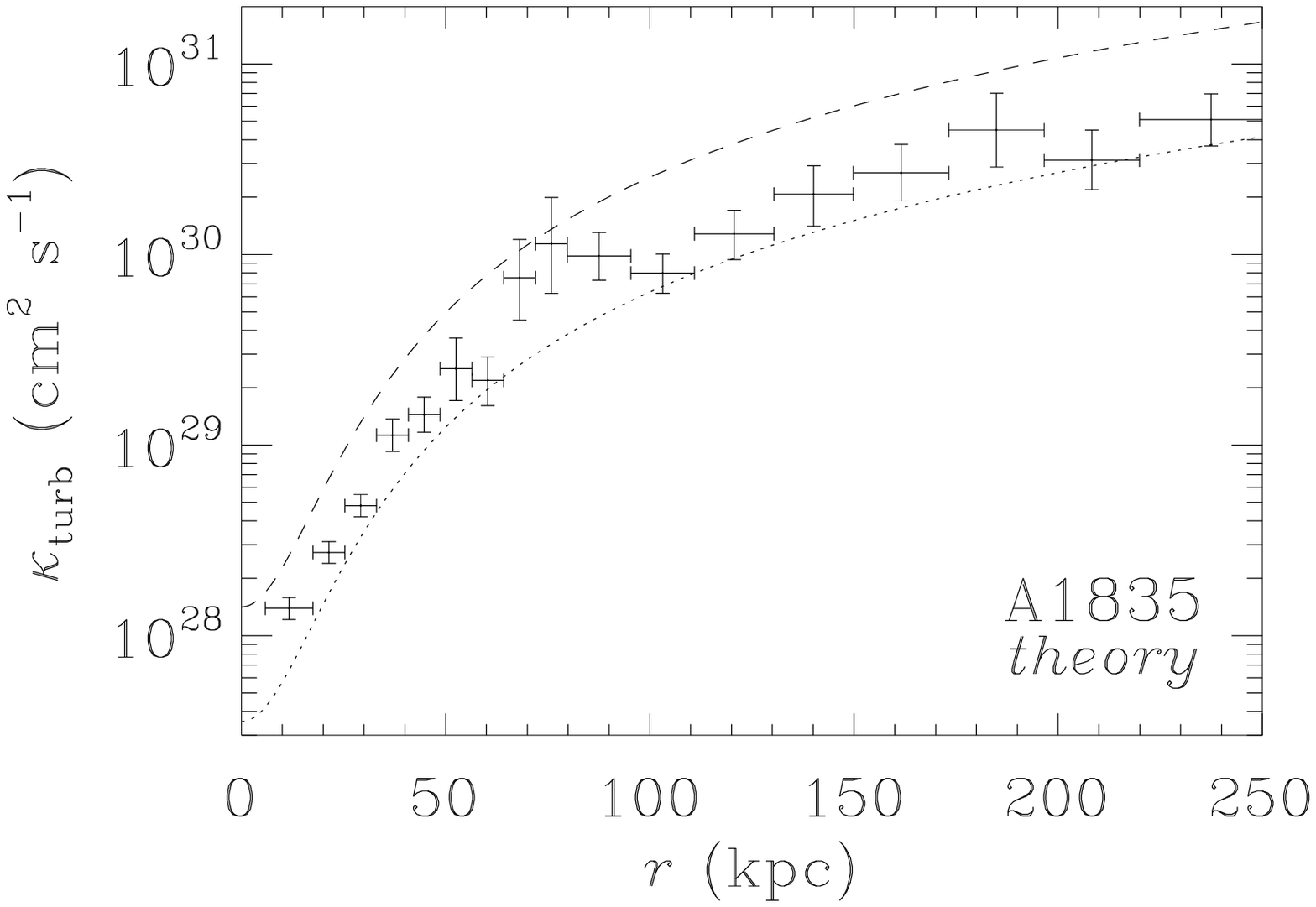}
\newline
\caption{({\it Left} to {\it right}, {\it top} to {\it bottom}) Profiles of the observed (deprojected) electron number density $n_{\rm e}$, observed (deprojected) temperature $T$, predicted magnetic-field strength $B$, predicted rms turbulent velocity $U_{\rm rms}$, predicted turbulence lengthscale $L$ and predicted turbulent diffusion coefficient $\kappa_{\rm turb}$ for the cool-core cluster A1835. The predicted data points correspond to $|\xi | = 0.75$, halfway between the firehose instability threshold ($|\xi |=1$; dotted line) and the mirror instability threshold ($|\xi | = 0.5$; dashed line). The line plots are derived from best-fitting analytic profiles to the electron number density and temperature (see footnote \ref{foot:profile}). The thick arrow on the plot of $U_{\rm rms}$ denotes the $182~{\rm km~s}^{-1}$ observationally-derived upper limit obtained by \citet{sfsp10} at $r\simeq 30~{\rm kpc}$.}
\label{fig:a1835}
\end{figure*}

\subsection{Turbulence scales}\label{sec:scales}

Implicit in the above discussion is the requirement that there be enough turbulent energy for the viscous heating rate mandated by the marginal stability condition (see eq. \ref{eqn:heating3}) to be maintained. We assume that throughout the cluster core all of the power from external stirring (see discussion in Section \ref{sec:discussion}) accepted by the turbulence is locally dissipated and thermalised via parallel viscosity:
\begin{equation}\label{eqn:dissipation}
m_{\rm i} n_{\rm i} \frac{U^2_{\rm rms}}{\tau_{\rm turb}} \simeq Q^+ ,
\end{equation}
where $\tau^{-1}_{\rm turb}$ is the effective rate at which energy is converted into turbulent motions. In conventional turbulence, this is equivalent to the eddy turnover timescale. In writing equation (\ref{eqn:dissipation}), we have ignored the possibility that energy could cascade to collisionless scales via Alfv\'{e}nic turbulence and heat the plasma via microphysical dissipation at the ion and electron Larmor scales \citep[see][and references therein]{scdhhqt09}.

Equations (\ref{eqn:Uprofile}) and (\ref{eqn:dissipation}) give
\begin{equation}\label{eqn:tauprofile}
\tau_{\rm turb} \simeq 2~ \xi^{-1} \left(\frac{n_{\rm e}}{0.1~{\rm cm}^{-3}}\right)^{-1} \left(\frac{T}{2~{\rm keV}}\right)~{\rm Myr} .
\end{equation}
We also define the characteristic turbulence lengthscale (`outer scale')
\begin{eqnarray}\label{eqn:Lprofile}
L &\equiv & U_{\rm rms}~\tau_{\rm turb} \nonumber\\*
\mbox{}&\simeq &0.2 ~\xi^{-3/2}\left(\frac{n_{\rm e}}{0.1~{\rm cm}^{-3}}\right)^{-1}\left(\frac{T}{2~{\rm keV}}\right)^{7/4}~{\rm kpc} .
\end{eqnarray}Another corollary of equations (\ref{eqn:Uprofile}) and (\ref{eqn:tauprofile}) is that the turbulent diffusion coefficient is
\begin{eqnarray}\label{eqn:kprofile}
\kappa_{\rm turb} &\sim& U^2_{\rm rms} \tau_{\rm turb} \nonumber\\*
\mbox{}&\simeq& 3\times 10^{27} ~\xi^{-2}\left(\frac{n_{\rm e}}{0.1~{\rm cm}^{-3}}\right)^{-1}\left(\frac{T}{2~{\rm keV}}\right)^{5/2}~{\rm cm}^2~{\rm s}^{-1} .\nonumber\\*
\end{eqnarray}
Note that $\kappa_{\rm turb}$ has the same scaling as the \citet{spitzer62} electron thermal diffusion coefficient (see Section \ref{sec:conduction}). Equations (\ref{eqn:Uprofile}) and (\ref{eqn:Lprofile}) further imply an effective Reynolds number associated with the parallel viscosity
\begin{equation}
{\rm Re} \equiv \frac{U_{\rm rms}L}{\kappa_{\rm visc}} = \frac{U_{\rm rms}L}{0.96(v^2_{\rm th}/\nu_{\rm ii})} = \frac{3}{\xi^2} ;
\end{equation}
($\kappa_{\rm visc}$ is the viscosity coefficient; see \citealt{braginskii65}). Thus, ${\rm Re}\sim 1$ -- $10$ and is independent of radius. In other words, the outer and viscous scales are close to one another, so that the motions are dissipated near the outer scale and there is no inertial range \citep[c.f.][]{fsccgw03}. However, we caution that an effective Reynolds number of order unity does not imply laminar flow in this case, since the turbulence is randomly stirred. We also stress that this is ${\rm Re}$ calculated on the basis of parallel collisional viscosity -- it does not imply a viscous cutoff for all plasma fluctuations, although, as stated above, we have ignored those as a possible heating channel.

\subsection{Cool-core cluster profiles: A1835}\label{sec:a1835}

Using equations (\ref{eqn:Bprofile}) -- (\ref{eqn:kprofile}), profiles of $B$, $U_{\rm rms}$, $L$ and $\kappa_{\rm turb}$ may be calculated for any given cluster. Here we provide results for one particular cool-core cluster, A1835 (we have performed the same exercise for many other clusters, with similarly sensible outcomes). In Fig.~\ref{fig:a1835}, we give the observed (deprojected) electron number density $n_{\rm e}$ and the observed (deprojected) temperature $T$, which are then used to predict the magnetic-field strength $B$, the rms turbulent velocity $U_{\rm rms}$, the characteristic turbulence scale $L$ and the turbulent diffusion coefficient $\kappa_{\rm turb}$. The density and temperature profiles are from \citet{sfsp10}. The data points for the predicted quantities use $|\xi | = 0.75$, but we also show upper and lower limits calculated by setting $|\xi | = 0.5$ (dashed lines) and $|\xi | = 1$ (dotted lines), respectively.

Equations (\ref{eqn:Bprofile}) -- (\ref{eqn:kprofile}) do not imply a specific causal relationship between the five quantities $n_{\rm e}$, $T$, $B$, $U_{\rm rms}$ and $L$. Given any two quantities, our theory can predict the other three. For example, if observations determined the profiles of $n_{\rm e}$ and $U_{\rm rms}$ for a given cluster, rather than of $n_{\rm e}$ and $T$, then our theory would predict $T$, $B$ and $L$. Ideally, our theory would be put to the test if three or more of these profiles were known observationally.

The predicted value for $B$ near the centre of the core is $\simeq 15$ -- $21~\mu{\rm G}$, decreasing to $\simeq 5$ -- $7~\mu{\rm G}$ at the outer core boundary. To our knowledge, observational estimates of the magnetic-field strength in A1835 have not yet appeared in the literature. As a radio mini-halo has recently been detected in A1835 \citep{mgmfgtc09}, there is hope for a magnetic field measurement there, although this might be quite a difficult task for such a distant cluster. The predicted turbulent velocity dispersion $U_{\rm rms}\sim 70$ -- $270~{\rm km~s}^{-1}$ throughout the core, attaining a value $U_{\rm rms}\simeq 114$ -- $162~{\rm km~s}^{-1}$ at a radius of $30~{\rm kpc}$. This is within the $182~{\rm km~s}^{-1}$ upper limit obtained by \citet{sfsp10} by measuring emission lines within $r\simeq 30~{\rm kpc}$, which is denoted on the plot by a thick arrow. A more conservative observational estimate of $U_{\rm rms}\lesssim 274~{\rm km~s}^{-1}$ was derived by treating the cluster as a point source and not applying any spatial smoothing \citep{sfsp10}. The predicted turbulence scale is $L\sim 0.2$ -- $0.7~{\rm kpc}$ near the centre of the core, increasing outwards to $\sim 70$ -- $200~{\rm kpc}$ near the temperature maximum. The predicted diffusion coefficient $\kappa_{\rm turb}$ rises sharply from $\sim 10^{28}~{\rm cm^2~s}^{-1}$ near the core centre to $\sim 10^{31}~{\rm cm^2~s}^{-1}$ at the outer core boundary. Given the uncertainties discussed in Sections \ref{sec:velocities} and \ref{sec:scales}, it is encouraging that these values are comparable to the inferred diffusion coefficients $\sim 10^{29}$ -- $10^{30}~{\rm cm}^2~{\rm s}^{-1}$ in a variety of observed clusters \citep{rcbf05,rcbf06,rcsbf08,dn08}.

\subsection{Non-cool-core cluster profiles}\label{sec:coma}

The plasma microphysics responsible for parallel viscous heating is in no way unique to cool-core clusters. Since plenty of turbulence is expected to be present (stirred by mergers, etc.), pressure anisotropies will develop and will presumably be maintained at a marginal level as described in Section \ref{sec:heat}. While heating and cooling times are quite long in such clusters and the situation can be quite time-dependent, we would nevertheless like to explore what the conjecture of an approximate local heating-cooling balance would imply, with the caveat that such a balance may in principle take a very long time to be established. We will see that such a balance leads to quite reasonable predictions that seem to be borne out by observational data. 

In isothermal clusters, equations (\ref{eqn:Bprofile}) -- (\ref{eqn:Lprofile}) imply that 
\begin{equation}
B\propto n_{\rm e}^{1/2} ,
\end{equation}
\begin{equation}
U_{\rm rms}\simeq {\rm const},
\end{equation}
\begin{equation}
\tau_{\rm turb}\propto L\propto n_{\rm e}^{-1}. 
\end{equation}
The scaling $B\propto n_{\rm e}^{1/2}$ in non-cool-core clusters has in fact already been observationally inferred for A2382 \citep{gmgpgrcf08}, Coma \citep{bfmggddt10} and A665 \citep{vmgfgob10}. 

The central density $n_{\rm e,c}\simeq 5\times 10^{-3}~{\rm cm}^{-3}$ and temperature $T\simeq 2.9~{\rm keV}$ in A2382 \citep{evb96} implies a thermal-equilibrium magnetic-field strength $B_{\rm c}\simeq 3.1~\xi^{-1/2}~\mu{\rm G}$, in excellent agreement with the $\sim 3~\mu{\rm G}$ estimate derived from rotation measure observations \citep{gmgpgrcf08}. For A2255, $n_{\rm e,c}\simeq 2\times 10^{-3}~{\rm cm}^{-3}$ \citep{fbgn97,gmfgdt06} and $T\simeq 3.5~{\rm keV}$ \citep[however, see \citealt{sp06}, who find temperature variations across A2255 from $T\sim 5.5$ -- $8.5~{\rm keV}$]{dw98}, the thermal-equilibrium magnetic-field strength $B_{\rm c}\simeq 2.2~\xi^{-1/2}~\mu{\rm G}$. This compares favourably with the observational estimate $B\sim 2.5~\mu{\rm G}$ obtained by \citet{gmfgdt06}, who used $B\propto n_{\rm e}^{1/2}$ in their analysis. In Table \ref{tab:bfields}, we list these and other central magnetic-field strength predictions.

Very recently there have been observational estimates of the magnetic-field strength profile in the non-cool-core Coma cluster by \citet{bfmggddt10}. They found that the best-fitting profile of the magnetic-field strength is
\begin{equation}\label{eqn:bcoma}
B_{\rm obs} \approx 4.7 \left(\frac{n_{\rm e}}{3.44\times 10^{-3}~{\rm cm}^{-3}}\right)^{1/2}~\mu{\rm G} ,
\end{equation}
with a `$\beta$-model' \citep{cf76} electron density profile
\begin{equation}\label{eqn:nefit}
n_{\rm e} = n_0 \left(1+\frac{r^2}{r^2_{\rm c}}\right)^{-3\beta/2} ,
\end{equation}
where $n_0 = 3.44 \times 10^{-3}~{\rm cm}^{-3}$, $r_{\rm c} = 291~{\rm kpc}$ and $\beta = 0.75$. Taking Coma to be an isothermal cluster with temperature $8.2~{\rm keV}$ \citep{arnaud01}, we find that the implied parallel viscous heating rate is (from eq. \ref{eqn:heating3} using eq. \ref{eqn:bcoma})
\begin{equation}
Q^+ \approx 1.6\times 10^{-28} \left(\frac{n_{\rm e}}{3.44\times 10^{-3}~{\rm cm}^{-3}}\right)^{2}~{\rm erg~s}^{-1}~{\rm cm}^{-3} .
\end{equation}
By way of comparison, the implied Bremsstrahlung cooling rate is (from eq. \ref{eqn:cooling2})
\begin{equation}\label{eqn:comacooling}
Q^- = 2.5\times 10^{-28} \left(\frac{n_{\rm e}}{3.44\times 10^{-3}~{\rm cm}^{-3}}\right)^{2}~{\rm erg~s}^{-1}~{\rm cm}^{-3} ,
\end{equation}
where $n_{\rm e}$ is given by equation (\ref{eqn:nefit}). While it is rather curious that these two rates are not only comparable, but also have the same radial scaling, we are not on solid ground applying our estimates to Coma because the cooling time for it implied by equation (\ref{eqn:comacooling}) is $t_{\rm cool} \sim n T / Q^- \sim$ the Hubble time at the cluster centre, increasing outwards in radius. The other isothermal clusters mentioned above and in Table \ref{tab:bfields} have relatively shorter cooling times, so one could in principle have a heating-cooling balance -- but no radial dependence of the magnetic-field strength has so far been measured.

\subsection{Postscript: thermal conduction}\label{sec:conduction}

\begin{figure*}
\centering
\includegraphics[width=3.2in]{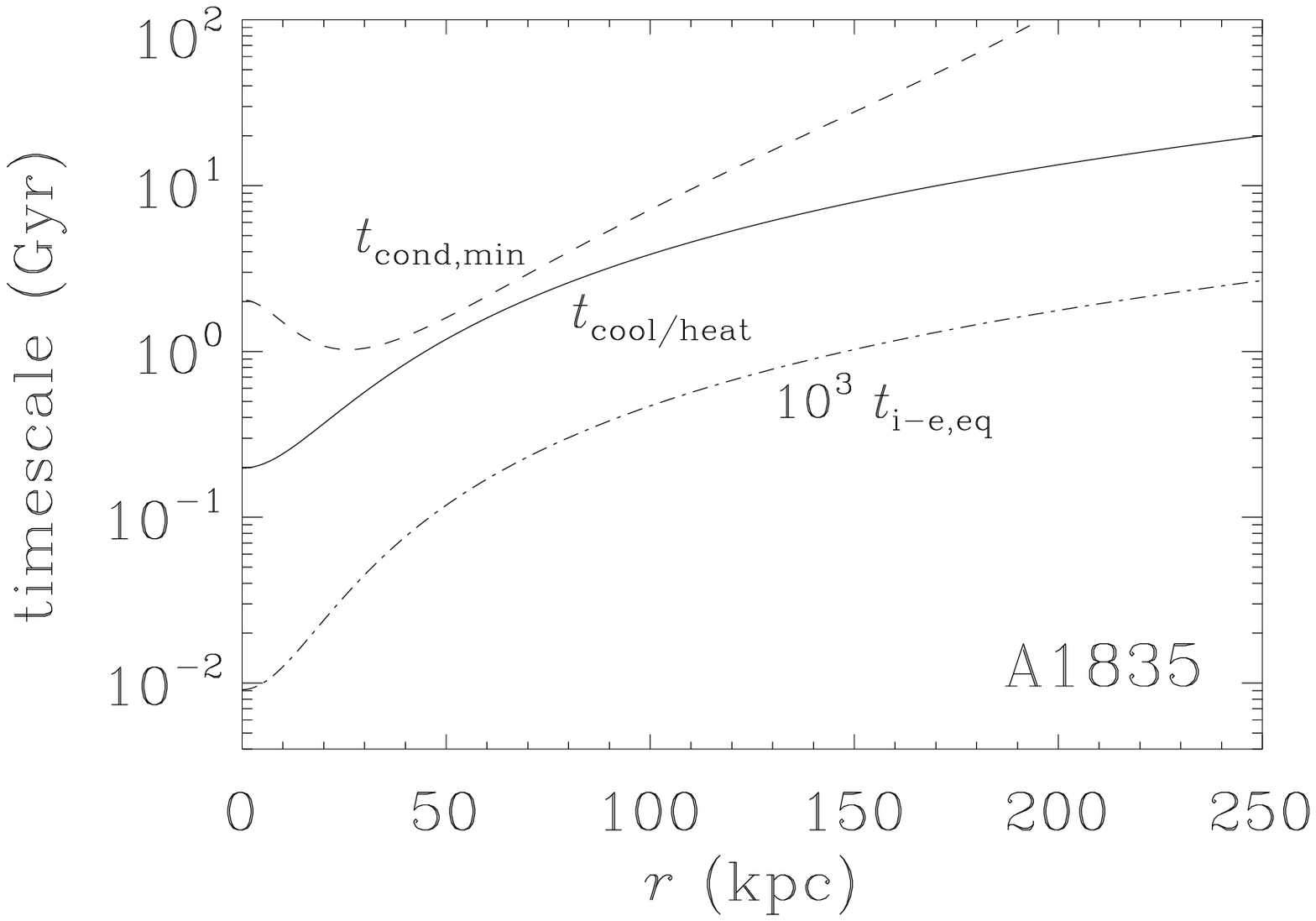}
\qquad
\includegraphics[width=3.2in]{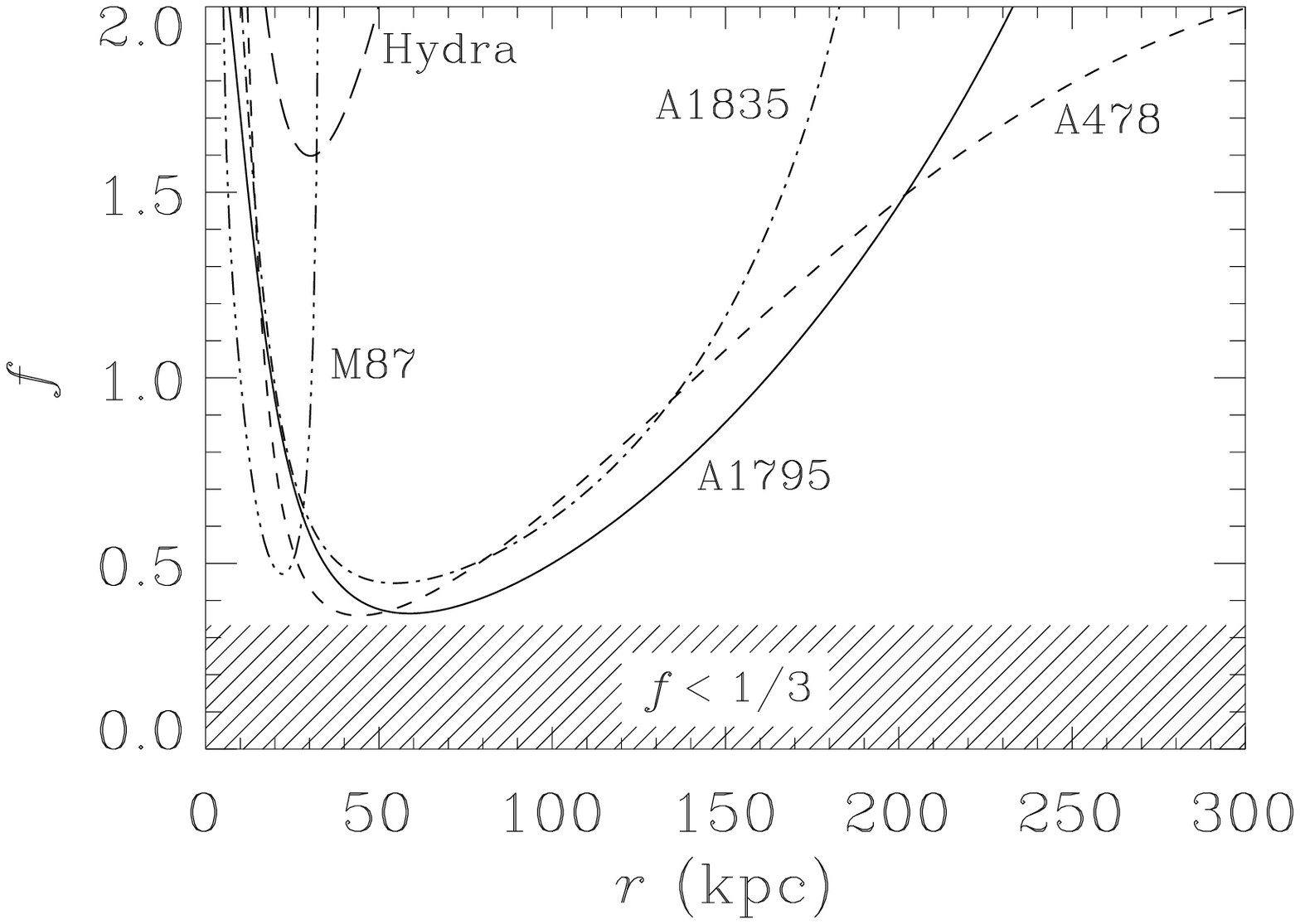}
\newline
\caption{(a) The cooling/heating timescale $t_{\rm cool/heat}$ (solid line), ion-electron equilibration timescale $t_{\rm i-e,eq}$ multiplied by $10^3$ (dash-dot line) and the shortest possible conduction timescale $t_{\rm cond,min}$ (dashed line) for A1835. (b) The Spitzer conduction suppression factor $f\equiv \kappa/\kappa_{\rm Sp}$ required for thermal conduction to balance radiative cooling as a function of radius for several clusters. Only suppression factors $f\lesssim 1/3$ (shaded region) are allowed in the presence of a tangled magnetic field. The line plots are derived from best-fitting analytic profiles to the electron number densities and temperatures (see footnote \ref{foot:profile}).}
\label{fig:conduction}
\end{figure*}

Bearing in mind that thermal conduction models routinely fail in the innermost regions of cool cluster cores \citep[e.g.][]{zn03,markevitch03,kaastra04,gmpd04}, it is important to note that parallel viscous heating should be especially important in these relatively cold ($T\sim 1~{\rm keV}$) and strongly-magnetised ($B\sim 10~\mu{\rm G}$) regions. A balance between parallel viscous heating and radiative cooling, however, does contain the implicit assumption that the thermal conduction is relatively unimportant. This ought to be checked.

The \citet{spitzer62} electron thermal diffusion coefficient is
\begin{equation}\label{eqn:ksp1}
\kappa_{\rm Sp} \simeq 10^{29} \left(\frac{n_{\rm e}}{0.1~{\rm cm}^{-3}}\right)^{-1} \left(\frac{T}{2~{\rm keV}}\right)^{5/2}~{\rm cm}^2~{\rm s}^{-1} .
\end{equation}
In tangled turbulent magnetic fields, the true thermal conductivity is expected to be $\kappa = f\kappa_{\rm Sp}$, where $f$ is the suppression factor, which is expected to range anywhere between $\sim 10^{-3}$ and $\sim 0.3$ \citep{tribble89,tao95,ps96,cc98,mk01,nm01,gruzinov02,clhkkm03}. Note that the Spitzer diffusion coefficient has the same density and temperature scaling as the turbulent diffusion coefficient derived in Section \ref{sec:scales}. We see that for $f\lesssim 0.03~\xi^{-2}$, turbulent heat diffusion is likely to be at least as important as the collisional conductivity.

In Fig. \ref{fig:conduction}a, we plot for A1835 the cooling/heating timescale
\begin{eqnarray}
\lefteqn{t_{\rm cool/heat} = \frac{3}{2}\frac{nT}{Q^-} = 0.21 \left(\frac{n_{\rm e}}{0.1~{\rm cm}^{-3}}\right)^{-1} \left(\frac{T}{2~{\rm keV}}\right)^{1/2}~{\rm Gyr} ,}\nonumber\\*\mbox{}
\end{eqnarray}
the ion-electron temperature equilibration timescale
\begin{equation}
t_{\rm i-e,eq} = 9.3\left(\frac{n_{\rm e}}{0.1~{\rm cm}^{-3}}\right)^{-1} \left(\frac{T}{2~{\rm keV}}\right)^{3/2}~{\rm kyr} ,
\end{equation}
and the shortest possible conduction timescale
\begin{equation}
t_{\rm cond,min} = \frac{3}{2} n T \left[ \frac{1}{r^2}\D{r}{}\left(f n_{\rm e}\kappa_{\rm Sp} r^2 \D{r}{T}\right)\right]^{-1}
\end{equation}
that one can expect in the presence of a tangled magnetic field (with $f = 1/3$; see, e.g., \citealt{gruzinov02}). Heating via thermal conduction is everywhere unimportant relative to radiative cooling and (by construction) parallel viscous heating. Also note that $t_{\rm i-e,eq} \lesssim 1~{\rm Myr}$ everywhere in the cluster core, so that the assumption of equal ion and electron temperatures is well justified.

These conclusions are by no means unique to A1835. To demonstrate this point, we plot in Fig. \ref{fig:conduction}b the Spitzer conduction suppression factor $f$ required for thermal conduction to balance radiative cooling as a function of radius for several different clusters.\footnote{\label{foot:profile}For the line plots in the figures, we have used analytic fits to the observed electron density and temperature profiles. One advantage is that this ensures smooth gradients for computing the conductive heating rates. The profiles for A478 and A1795 were taken from \citet{dc05} and the profiles for M87 were taken from \citet{gmpd04}. For A1835 \citep{sfsp10} and Hydra A \citep{dnmfjprw01}, the electron number density was fit using equation (\ref{eqn:nefit}) with $\beta=0.593$, $r_{\rm c}=32.42~{\rm kpc}$ and $n_0=0.115~{\rm cm}^{-3}$ (for A1835) and $\beta=0.393$, $r_{\rm c}=10.9~{\rm kpc}$ and $n_0=0.0669~{\rm cm}^{-3}$ (for Hydra A). Their temperature profiles were fit using equation (22) of \citet{dc05} with $T_0=9.55~{\rm keV}$, $T_1=7.39~{\rm keV}$, $r_{\rm ct}=33.35~{\rm kpc}$ and $\delta=0.62$ (for A1835) and $T_0=3.95~{\rm keV}$, $T_1=0.856~{\rm keV}$, $r_{\rm ct}=30.60~{\rm kpc}$ and $\delta=0.48$ (for Hydra A).} Only suppression factors $f\lesssim 1/3$ are expected in the presence of a tangled magnetic field. None of these clusters can be in thermal balance between thermal conduction and radiative cooling.\footnote{Note that there do exist some clusters for which a fraction of Spitzer conductivity may be able to stabilise their cores against a cooling catastrophe (e.g. \citealt{zn03}; see however \S~3.3 of \citealt{co08} for an alternative point of view.)}

\section{Discussion}\label{sec:discussion}

\begin{figure}
\centering
\includegraphics[width=3.2in]{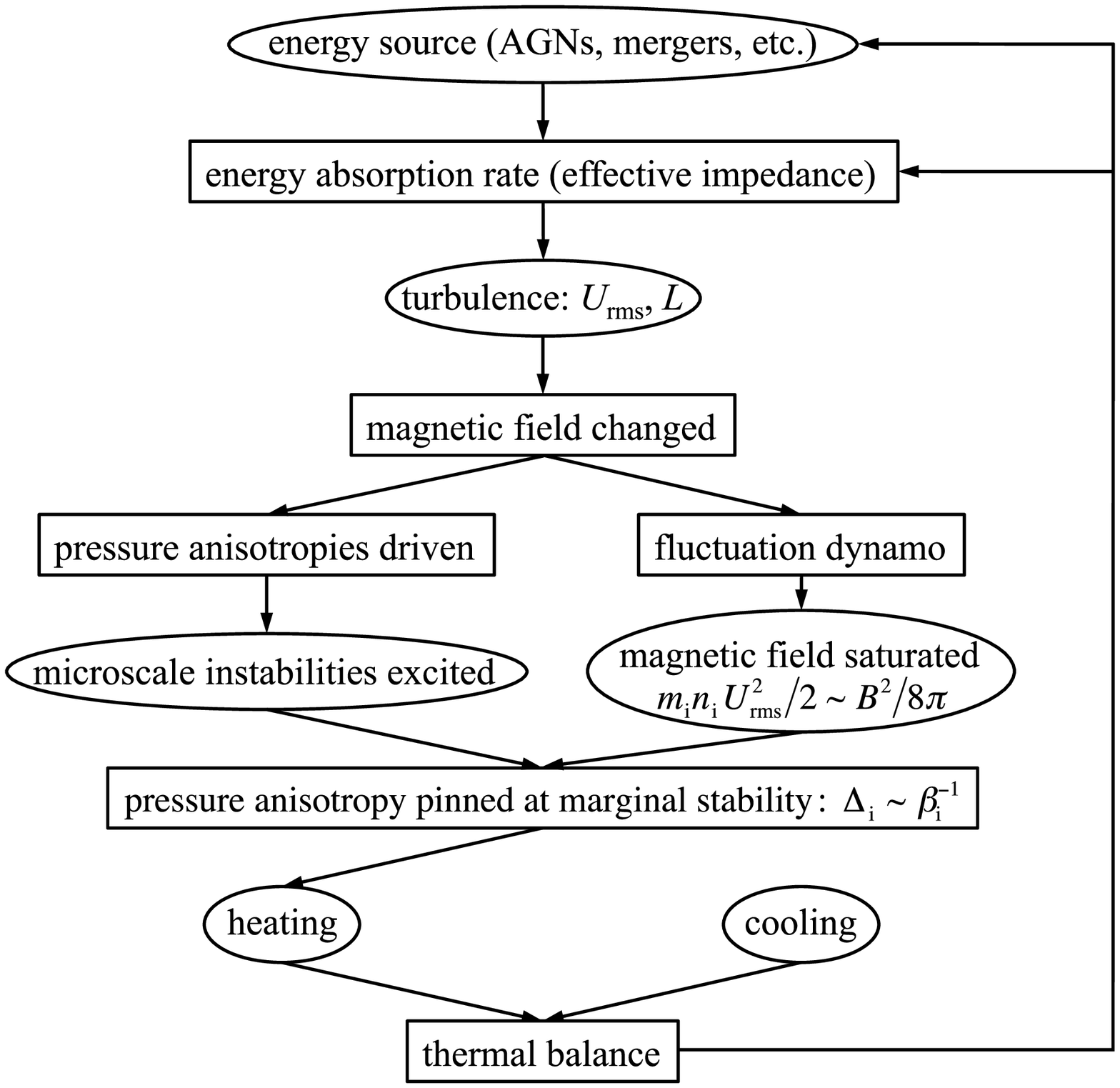}
\caption{Qualitative outline of the processes responsible for thermal balance between parallel viscous heating and radiative cooling. Energy sources (AGNs, mergers, etc.) inject energy into the ICM, some or all of which (depending on the effective impedance of the ICM) is absorbed by the plasma and converted into turbulence. This changes the magnetic-field strength, giving rise to both a fluctuation dynamo and pressure anisotropies. The fluctuation dynamo presumably saturates in equipartition between the turbulent magnetic and kinetic energies. The pressure anisotropies excite microscale instabilities, whose effect is to pin the pressure anisotropy at marginal stability. The pressure anisotropy determines the viscous stress and therefore the heating rate. This heating balances radiative cooling, giving rise to a stable thermal equilibrium that is maintained by energy injection by self-regulated sources (such as AGNs) and/or self-regulated energy absorption by the turbulence.}
\label{fig:outline}
\end{figure}

In this paper, we have introduced a model for regulating cooling in cluster cores in which turbulence, magnetic fields and plasma physics all play crucial roles. Our findings are fundamentally based on an appreciation that, whatever the source of effective viscosity in the ICM, it is certainly not hydrodynamic. Instead, it is set by the microscale plasma-physical processes, which are inevitable in a weakly collisional, magnetised environment such as the ICM. Assuming that microscale plasma instabilities pin the pressure anisotropy at its marginally stable value, we derive an expression for the heating rate due to parallel viscous dissipation of turbulent motions. For typical conditions in a variety of cluster cores, this rate is of the right magnitude to balance radiative cooling. Moreover, this source of heating turns out to be thermally stable. Put simply, the viscous stress in the ICM is a dynamic quantity that responds to local changes in temperature, density and magnetic-field strength in such a way as to prevent runaway heating or cooling. A basic qualitative outline of how this occurs is given in Fig. \ref{fig:outline}. If true, what we have conjectured constitutes a physical mechanism that allows clusters to develop stable non-isothermal temperature profiles and avoid a cooling catastrophe -- an outcome that has remained elusive in models involving balancing the cooling by thermal conduction.

Of the five quantities -- density, temperature, rms magnetic-field strength, rms turbulent velocity, characteristic turbulence scale -- given radial profiles of any two, we can predict the rest. The most reliable, as well as readily observable, predictions that follow from our theory concern the (rms) magnetic-field strength. For typical electron densities and temperatures, we predict magnetic-field strengths in the range $\sim 1$ -- $10~\mu{\rm G}$, well within current observational constraints. For the specific clusters discussed here (e.g. A1835, Hydra A, A2199, A2382, A2255), a balance between parallel viscous heating and radiative cooling results in field strengths and profiles that are quite reasonable and in good agreement with current observational estimates where available. It would be interesting to test our model further via analysis of Faraday rotation maps.

Another prediction is the manner in which the magnetic-field strength $B$ depends on electron density $n_{\rm e}$ and temperature $T$ (eq. \ref{eqn:Bprofile}):
\begin{equation}
B\propto n^{1/2}_{\rm e} T^{3/4} \quad \textrm{(for cool-core clusters).}
\end{equation}
In cool-core clusters, this suggests that the oft-employed assumption of an exclusive relationship between $B$ and $n_{\rm e}$ should be relaxed. It is encouraging that, in isothermal clusters, the implied scaling
\begin{equation}\label{eqn:bscalingiso}
B\propto n^{1/2}_{\rm e} \quad \textrm{(for isothermal clusters)}
\end{equation}
has already been observationally inferred to hold in A2382 \citep{gmgpgrcf08}, Coma \citep{bfmggddt10} and A665 \citep{vmgfgob10}. A caveat here is that a local heating-cooling balance in such clusters may take very long to establish and so should be treated with due scepticism. A test of whether this conjecture is reasonable would be to determine whether equation (\ref{eqn:bscalingiso}) is satisfied systematically in a large sample of isothermal clusters.

There are several peripheral consequences of our predictions that warrant further discussion. Firstly, the relatively large magnetic-field strengths inferred from observed rotation measures and predicted by a balance between parallel viscous heating and radiative cooling are stable or at least marginally stable to the HBI. The stability criterion for the HBI \citep{quataert08} may be written as a lower-bound on the magnetic-field strength:
\begin{eqnarray}\label{eqn:hbi}
\lefteqn{B \gtrsim 3\left(\frac{g}{10^{-8}~{\rm cm~s}^{-2}}\right)^{1/2} \left(\frac{n_{\rm e}}{0.1~{\rm cm}^{-3}}\right)^{1/2}\left(\frac{r}{100~{\rm kpc}}\right)^{1/2} }\nonumber\\*\mbox{} &&\times \left(\frac{{\rm d}\ln T/{\rm d}\ln r}{0.1}\right)^{1/2}~\mu{\rm G} . 
\end{eqnarray}
Even under the favourable conditions we have chosen for the electron density $n_{\rm e}$, the radial distance from the cluster centre $r$ and the temperature slope $d\ln T/d\ln r$, the magnetic fields predicted in this paper exceed the stability limit (\ref{eqn:hbi}). Indeed, \citet[\S~3.3]{brbp09} and \citet[\S~5.7]{pqs09} found that simulated clusters with magnetic-field strengths $B\sim 3~\mu{\rm G}$ demonstrated a delayed cooling catastrophe due to the stabilisation of HBI modes by magnetic tension. By contrast, the rms magnetic-field strength used in the \citet{pqs10} simulation, where a combination of turbulent stirring and HBI-governed thermal conduction gave a cool core in long-term thermal balance, was $10^{-9}~{\rm G}$. More observational estimates of magnetic-field strengths in a variety of clusters would clearly be very useful to help realistic modelling and testing of theories.

Secondly, the same process that is responsible for stably heating the ICM in our model may also influence the turbulent diffusion of metals throughout the cores of galaxy clusters. That our values for the turbulent diffusion coefficient $\kappa_{\rm turb}$ are comparable to those inferred in a variety of clusters is encouraging. However, \citet{rcbf05} found that diffusion coefficients that increase with radius (as ours does) imply abundance profiles that are too centrally peaked compared to those observed. It is therefore worth investigating the diffusion of metals using our predicted scaling $\kappa_{\rm turb} \propto n^{-1}_{\rm e} T^{5/2}$.

Thirdly, turbulent heat diffusion may become dominant relative to collisional electron heat conduction if the Spitzer conductivity suppression factor $f\lesssim 0.03$ (see eqns \ref{eqn:kprofile} and \ref{eqn:ksp1}). However, for typical density and temperature profiles of cool-core clusters, neither form of heat diffusion seems to be important relative to parallel viscous heating (see Section \ref{sec:conduction}).

It is prudent to reiterate here our assumptions and their limitations:
\begin{enumerate}
\item We have implicitly assumed that there is enough turbulent energy so that, if it is thermalised via parallel viscous heating, it can offset cooling. This requires either the source of the turbulent energy or the amount of energy locally accepted by the plasma and converted into turbulent motions to have some knowledge of the cooling rate and to be self-regulating. 

AGNs are a natural candidate for providing a self-regulating energy source, whether the stirring is due to AGN-driven jets, bubbles, weak shocks/sound waves, gravitational modes and/or cosmic-ray--buoyancy instabilities. Observationally, a large majority of cool-core clusters harbour radio sources at their centres, and the AGN energy output inferred from radio-emitting plasma outflows and cavities is often similar to the X-ray cooling rate of the central gas \citep[e.g.][]{bcsrm10}.

However, it may not be necessary for the energy source to be fine-tuned or tightly self-regulating. Our scheme for heating the ICM can deal with excess turbulent energy via the following two considerations. First, not all of the power provided by the external stirring has to be locally thermalised via turbulence, as the turbulence may have an effective `impedance' and only accept the amount of power that can be locally viscously dissipated without triggering the microinstabilities; the remaining power could be transported elsewhere. Second, the parallel viscous heating rate (eq. \ref{eqn:heating2}), coupled with the assumption of equipartition kinetic and magnetic energies (eq. \ref{eqn:equipartition}), is naturally self-regulating: any increase in turbulent energy implies an increase in magnetic energy, which implies an increased viscous dissipation rate. Both of these regulation mechanisms require there to be a sufficient amount of turbulence to pin the pressure anisotropy at its marginal stability threshold. As it happens, there seems to be no dearth of turbulent power \citep[e.g.][]{cfjsb04}. The fact that some of this power can be thermalised via parallel viscous dissipation in a thermally stable way provides an attractive alternative to other heating models that suffer from stability issues. Further work on the excitation and replenishment of turbulence in a weakly collisional ICM is clearly needed.
\item The heating is assumed to be all due to parallel collisional viscosity. In principle, energy could cascade through the parallel viscous scale and onwards to collisionless scales via Alfv\'{e}nic turbulence and then heat the plasma via microphysical dissipation at the ion and electron Larmor scales \citep[see][and references therein]{scdhhqt09}. This possibility has been ignored here.
\item The predicted turbulent velocity, obtained by assuming approximate equipartition between kinetic and magnetic energies, is probably a lower limit on the actual turbulent velocity since $U_{\rm rms}$ is unlikely to be smaller than the Alfv\'{e}n speed. Unfortunately, a more exact estimate of $U_{\rm rms}$ would require a detailed understanding of turbulent dynamo saturation in the ICM, far beyond the scope of the present work.
\item Our estimates of the characteristic turbulence scale and turbulent diffusion coefficient are perhaps the more uncertain of our predictions, as they depend on the rate of transfer of energy from the external driving sources into the turbulence. This encodes what we have referred to above as the effective `impedance' of the turbulence, the detailed physics of which is poorly understood.
\end{enumerate}
All these concerns clearly require further theoretical work. However, the fact that our predictions are not too far from current observational estimates or plausible expectations lends us hope that these concerns might be of secondary importance.

While the exact numbers predicted in this paper should be taken with a grain of salt, one cannot help being encouraged by the fact that they seem to be quite reasonable without any fine-tuning of adjustable prefactors. The basic conjecture that the saturation of microscale plasma instabilities endows the ICM with a thermally stable source of viscous heating seems robust. If our predictions are confirmed, this would constitute strong evidence that microphysical plasma processes play a decisive role in setting the large-scale structure and evolution of galaxy clusters.

\section*{Acknowledgments}

It is a pleasure to thank Steve Balbus, Torsten En\ss lin, Andy Fabian, Petr Kuchar, Ian Parrish, Eliot Quataert, Mark Rosin and Prateek Sharma for useful discussions, Helen Russell and Annalisa Bonafede for graciously providing observational data for several clusters, and an anonymous referee for constructive comments. This work was supported by the STFC.

\bsp
\label{lastpage}

\end{document}